\documentclass[twoside,twocolumn,10pt]{article}
\usepackage{tabularx}
\usepackage{extsizes}
\usepackage[super,sort&compress,comma]{natbib} 
\usepackage[version=3]{mhchem}
\usepackage[left=1.5cm, right=1.5cm, top=1.785cm, bottom=2.0cm]{geometry}
\usepackage{balance}
\usepackage{mathptmx}
\usepackage{amsmath}
\usepackage{amssymb}
\usepackage{amsbsy}
\usepackage{siunitx}
\usepackage{amsfonts}
\usepackage{sectsty}
\usepackage{graphicx} 
\usepackage{lastpage}
\usepackage[format=plain,justification=justified,singlelinecheck=false,font={stretch=1.125,small,sf},labelfont=bf,labelsep=space]{caption}
\usepackage{float}
\usepackage{fancyhdr}
\usepackage{fnpos}
\usepackage[english]{babel}
\addto{\captionsenglish}{%
  
}
\usepackage{array}
\usepackage{booktabs}
\usepackage{droidsans}
\usepackage{charter}
\usepackage[T1]{fontenc}
\usepackage[usenames,dvipsnames]{xcolor}
\usepackage{setspace}
\usepackage[compact]{titlesec}
\usepackage{hyperref}

\usepackage{epstopdf}

\definecolor{cream}{RGB}{222,217,201}

\begin{document}

\pagestyle{fancy}
\thispagestyle{plain}
\fancypagestyle{plain}{
\renewcommand{\headrulewidth}{0pt}
}

\makeFNbottom
\makeatletter
\renewcommand\LARGE{\@setfontsize\LARGE{15pt}{17}}
\renewcommand\Large{\@setfontsize\Large{12pt}{14}}
\renewcommand\large{\@setfontsize\large{10pt}{12}}
\renewcommand\footnotesize{\@setfontsize\footnotesize{7pt}{10}}
\makeatother

\renewcommand{\thefootnote}{\fnsymbol{footnote}}
\renewcommand\footnoterule{\vspace*{1pt}%
\color{cream}\hrule width 3.5in height 0.4pt \color{black}\vspace*{5pt}} 
\setcounter{secnumdepth}{5}

\makeatletter 
\renewcommand\@biblabel[1]{#1}            
\renewcommand\@makefntext[1]%
{\noindent\makebox[0pt][r]{\@thefnmark\,}#1}
\makeatother 
\renewcommand{\figurename}{\small{Fig.}~}
\sectionfont{\sffamily\Large}
\subsectionfont{\normalsize}
\subsubsectionfont{\bf}
\setstretch{1.125} 
\setlength{\skip\footins}{0.8cm}
\setlength{\footnotesep}{0.25cm}
\setlength{\jot}{10pt}
\titlespacing*{\section}{0pt}{4pt}{4pt}
\titlespacing*{\subsection}{0pt}{15pt}{1pt}

\newcommand\bs{\emph{B. subtilis}} 
\newcommand\ec{\emph{E. coli}} 
\newcommand\pa{\emph{P. aeruginosa}} 
\newcommand\ugml{$\mu$g/ml} 
\newcommand\nvec{\mathbf{n}}

\newcommand{\matriz}[1]{\boldsymbol{\mathsf{#1}}}
\newcommand{\unitvc}[1]{\hat{\mathbf{#1}}}
\newcommand{\Qvec}{\boldsymbol{\mathsf{Q}}}
\newcommand{\uvec}{\mathbf{u}}

\fancyhead{}
\renewcommand{\headrulewidth}{0pt} 
\renewcommand{\footrulewidth}{0pt}
\setlength{\arrayrulewidth}{1pt}
\setlength{\columnsep}{6.5mm}
\setlength\bibsep{1pt}

\makeatletter 
\newlength{\figrulesep} 
\setlength{\figrulesep}{0.5\textfloatsep} 

\newcommand{\topfigrule}{\vspace*{-1pt}%
\noindent{\color{cream}\rule[-\figrulesep]{\columnwidth}{1.5pt}} }

\newcommand{\botfigrule}{\vspace*{-2pt}%
\noindent{\color{cream}\rule[\figrulesep]{\columnwidth}{1.5pt}} }

\newcommand{\dblfigrule}{\vspace*{-1pt}%
\noindent{\color{cream}\rule[-\figrulesep]{\textwidth}{1.5pt}} }

\makeatother

\twocolumn[
  \begin{@twocolumnfalse}
\vspace{1em}
\sffamily
\begin{tabular}{m{4.5cm} p{13.5cm} }

    & \noindent\LARGE{\textbf{Ordering in Confined Two-Dimensional Nematic Systems: Mesoscopic Simulations Based on Different Mean-Field Potentials}} \\
\vspace{0.3cm} & \vspace{0.3cm} \\

 & \noindent\large{Humberto H\'{\i}jar\textit{$^{a}$} and  Apala Majumdar\textit{$^{b}$} } \\

    & \noindent\normalsize{We use nematic Multi-particle Collision Dynamics (N-MPCD) simulations to study confined nematic liquid crystals in square domains, with three distinct mean-field potentials: the classical Maier-Saupe and Marrucci-Greco models, and a more recent model due to Ilg, Karlin, and \"Ottinger. These potentials incorporate diverse physical features, including spatial gradients and nonlinear dependencies on the order parameter, to describe nematic ordering at mesoscopic scales. We derive coarse-grained equations from a Fokker–Planck description with tensorial closures, and analyse the emergence of order as a function of interaction strength, $U$, in two dimensions. The critical interaction strength depends on the choice of the mean-field potential. We also analytically estimate the nematic coherence length in three dimensions, to establish a rigorous correspondence between the N-MPCD parameters (the system size $R$ and $U$) and the continuum Landau-de Gennes theoretical parameters. We systematically study equilibrium and metastable configurations, including relaxation pathways to stable equilibria, on square domains, for all three mean-field potentials. Our results confirm universal equilibrium and metastable configurations for all three mean-field potentials. Our results also suggest that the N-MPCD predictions are consistent with the continuum Landau-de Gennes predictions, regardless of the choice of the underlying mean-field potential and approximations, for large $R$ and $U$. There are differences for small $R$ and for $U$ near the critical interaction strength, that need to be further explored and quantified for new-age multiscale and multiphysics theories.}
\end{tabular}

 \end{@twocolumnfalse} \vspace{0.6cm}

  ]

\renewcommand*\rmdefault{bch}\normalfont\upshape
\rmfamily
\section*{}
\vspace{-1cm}


\footnotetext{\textit{$^{a}$~Research Center, La Salle University Mexico, Benjamin Franklin 45, 06140, Mexico City, Mexico and Faculty of Sciences, National Autonomous University of Mexico, Circuito Exterior S/N, Ciudad Universitaria, 04510, Mexico City, Mexico}}
\footnotetext{\textit{$^{b}$~Department of Mathematics, University of Manchester, Manchester, M13 9PL, UK}}

\footnotetext{\dag~Electronic Supplementary Information (ESI) available. See DOI: 10.1039/cXsm00000x/}


%

\section{\label{introduction_section}Introduction}

Nematic liquid crystals (NLCs) are classical examples of partially ordered materials that combine fluidity with the ordering characteristics of conventional solids \cite{de_gennes_1993}. The constituent nematic molecules are typically elongated or rod-like, and these rod-like molecules translate freely and align along certain locally preferred directions, referred to as nematic directors. Consequently, NLCs have long-range orientational order, direction-dependent optical, mechanical and rheological properties and have been the working material of choice for the multi-billion dollar display industry, for many decades~\cite{lagerwallreview}. Contemporary NLC research has expanded into new territories such as healthcare technologies, metamaterials, robotics, energy applications etc. As such, it is of prime conceptual and potentially, practical importance to qualitatively and quantitatively understand how molecular properties and micromechanics can be accurately coarse-grained to deliver analytically tractable and computationally efficient continuum or macroscopic mathematical theories.

NLCs, like most complex materials, can be modelled at different length and time scales. There are molecular-level models for NLCs, such as lattice-based and off-lattice based models, that have explicit information about molecular shapes and intermolecular interactions. Such models are usually based on an inter-particle interaction energy and the equilibrium configurations are modelled by minimizers of the prescribed interaction energy; the energy minimizers are typically computed using Monte Carlo and Molecular Dynamics methods~\cite{robinson_liq_cryst_2017}. Such detailed approaches are computationally expensive and are not suitable for simulating modern NLC applications-oriented systems. Mean-field theories are mesoscopic theories wherein many-body interactions are averaged by a mean-field. Each nematic molecule interacts with this mean field and the total mean-field energy is the sum/integral of these interaction energies. The Maier-Saupe~(MS) theory is a celebrated mean-field theory for uniaxial nematics, composed of rod-like molecules, that can successfully predict the isotropic-nematic phase transition~\cite{maier_z_naturforschung_1959}. Mean-field theories have not been widely used for simulating large practical systems because the energy minimization problem has many implicit features, such as an unknown probability distribution function for the molecular orientations, etc. Continuum theories for NLCs are macroscopic theories, well suited for capturing observable NLC phenomena. Continuum theories for NLCs are based on the concept of a macroscopic order parameter, that contain information about the macroscopic state of NLC ordering - the directors and the averaged orientational ordering, that can be related to experimentally measurable quantities such as birefringence data or dielectric data~\cite{patranabish2021}. Continuum theories are variational phenomenological theories, based on an energy functional that depends on the macroscopic order parameter and its derivatives, such that the model parameters have no explicit connection with the underlying molecular details. 

There are at least three continuum theories for NLCs, based on the underpinning assumption of constituent rod-like nematic molecules~\cite{liu2001}. The Oseen-Frank theory is the simplest continuum theory restricted to uniaxial nematics, with a single director or a single distinguished direction of averaged molecular alignment such that all directions orthogonal to the director are physically equivalent, with a constant degree of orientational ordering. The Oseen-Frank theory cannot capture NLC systems with multiple directors (biaxial systems) and can only describe low-dimensional NLC defects as such. The Ericksen theory is a continuum theory for uniaxial nematics with variable order parameter; in other words, the Ericksen theory is based on the premise that order parameters vanish at nematic defects to accommodate the defect distortions and can accommodate higher-dimensional defects. The Landau-de Gennes (LdG) theory is a celebrated and powerful continuum theory for generic NLC phases, not restricted to the assumption of uniaxiality. The LdG theory describes the macroscopic NLC state by the a macroscopic order parameter, the $\Qvec$-order parameter which is a symmetric, traceless $3\times 3$ matrix relatable to the dielectric anisotropy or magnetic susceptibility tensor~\cite{de_gennes_1993}. The LdG free energy in three dimensions (3D) is typically of the form
\[
I_{\text{LdG}}[\Qvec]: = \int_{\mathcal{B}} f_{\text{B}}(\Qvec) + w(\Qvec, \nabla \Qvec)~dV
\]
where $\mathcal{B}$ is a 3D domain, $f_{\text{B}}$ is the LdG bulk energy
\[
f_{\text{B}}(\Qvec) = \frac{A}{2}|\Qvec|^2 - \frac{B}{3}\textrm{tr}\Qvec^3 + \frac{C}{4}\left(|\Qvec|^2 \right)^2,
\] where $A$ is the rescaled temperature, $B$ and $C$ are positive material-dependent constants. The bulk energy models the energy of spatially homogeneous NLC samples and predicts a first-order isotropic-nematic phase transition at $A = \frac{B^2}{27C}$. The elastic energy density, $w$, penalises spatial inhomogeneities and is typically a quadratic and convex function of the gradient of the order parameter, $\nabla \Qvec$, and involves material-dependent elastic constants. The bulk and elastic LdG parameters are determined by fitting with experimental data, and the bulk energy is simply a Taylor expansion about the isotropic state, $\Qvec=0$, near the isotropic-nematic phase transition. The physically observable configurations are modelled by LdG free energy minimizers subject to the imposed boundary conditions. The energy minimizers, and in fact all critical points of the LdG free energy, are analytic solutions of the associated Euler-Lagrange equations: a nonlinear system of five elliptic, coupled partial differential equations. There has been substantial recent analytic and numerical work on the critical points of the LdG free energy~\cite{han_siam_2020}. The LdG theory has been successful in capturing NLC phenomena (structural transitions, phase transitions, switching processes) but the LdG theory can fail at low temperatures, for small domains, near complex defects or effectively whenever small-scale or micro-effects are important.

There are important but partially open questions: what sort of liquid crystal systems are amenable to LdG-type approaches and can we quantify the limits of validity of such continuum phenomenological theories? These questions rely on a systematic understanding of the relationships between the different levels of modelling, \textit{i.e.}, when do molecular or mean-field predictions agree with LdG predictions and when do they differ, and if so, why? We focus on comparisons between mean-field and continuum LdG models in this paper. The mean-field models are studied by means of the N-MPCD (nematic Multi-particle Collision Dynamics) method~\cite{shendruk_soft_matter_2015}. N-MPCD is a particle-based stochastic algorithm for simulating the positions, velocities and orientations of nematic packets in cells, based on collision operators. The collision operators update the nematic particle orientations based on mean-field potentials and have embedded hydrodynamics, \textit{i.e.}, the two-way coupling between the nematic order and the fluid velocity. N-MPCD simulations respect frame indifference, material symmetries, conservation of mass, linear and angular momentum, and are more detailed than phenomenological LdG descriptions by definition.

In~\cite{hijar_soft_matter_2024}, the authors use the N-MPCD method with the mean-field MS potential to simulate NLC systems within regular 2D polygons, with tangent boundary conditions. The tangent boundary conditions require the nematic packet orientations to be tangent to the polygon edges. This problem was studied in depth in~\cite{han_siam_2020} in a reduced LdG framework. The authors conclude that MS and the LdG predictions are consistent for large polygons, or for relatively low temperatures that favour well-ordered systems. It is remarkable that the MS and the reduced LdG predictions are not only consistent for equilibrium and observable configurations, but also for metastable transient configurations, in the macroscopic well-ordered regime. The authors also compute defect annihilation laws relevant for relaxation dynamics or pathways to equilibrium and the finite-size effects can slow defects and attract defects to vertices.

In this paper, we use N-MPCD methods to simulate NLCs on square domains as a representative example, with tangent boundary conditions, and with three different mean-field potentials: the standard MS potential, the Marrucci–Greco (MG) potential often used in polymeric liquid crystal modelling~\cite{marrucci_mol_cryst_liq_cryst_1991}, and a mean-field potential introduced by Ilg, Karlin, and Öttinger (IKÖ), which accounts for entropy-consistent dynamics~\cite{ilg_phys_rev_e_1999}. All potentials are characterized by an interaction strength parameter, $U$, such that larger values of $U$ correspond to lower temperatures or well ordered systems that favour potentially uniaxial nematic ordering. Each model is based on the underpinning assumption of rod-shaped nematic molecules; the MG potential accounts for spatial inhomogeneities and can be better suited for frustrated systems whereas the IK\"O potential captures a nonlinear dependence of the interaction strength on the nematic ordering. We conduct detailed N-MPCD simulations of NLCs in square domains, over systems sizes (defined by $R$) spanning two orders of magnitude, and suitably defined intervals for $U$ that can capture the transition from weakly ordered to well-ordered regimes.

We undertake a detailed study of the mappings between the LdG parameters and the N-MPCD parameters---$U$ and $R$. In two dimensions (2D), we average the Fokker-Planck equation~(FPE) for the evolution of the probability distribution function (p.d.f) of the molecular orientations. We assume closure relations to relate higher-order order parameters to the standard scalar order parameter (related to the second moment of the p.d.f.) and use these closure relations to predict the critical interaction strength for the onset of nematic ordering in 2D, for all three mean-field potentials. This is a novelty of our work: model-specific critical interaction strengths, $U_{\text{c}}$. We then use a 3D closure scheme proposed in~\cite{kroeger_j_nonnewton_fluid_mech_2008} to estimate the nematic coherence length, as a function of the nematic elastic constant and the LdG bulk constants. The nematic coherence length is a material-dependent length scale, related to the nematic defect core sizes dictated by the competition between the elastic energy and the LdG bulk energy. The main purpose of the estimations of the nematic coherence length is to define a parameter, $\lambda$, that is the ratio of the geometric system size to the nematic coherence length, so that the N-MPCD simulations can be interpreted in the context of the (reduced) LdG results on nematic equilibria on square domains. Informally speaking, small $\lambda$ refers to nano-scale domains whereas large $\lambda$ corresponds to large experimentally relevant systems. Notably, $\lambda$ also contains information about the embedded MS potential and this method can be generalized to other mean-field potentials too. This is also a novelty of our work---a N-MPCD or mean-field analogue of the nematic coherence length (usually defined in terms of phenomenological LdG coefficients) and $\lambda$ in terms of $R$ and $U$. We find that the N-MPCD predictions are wholly consistent with each other for all three mean-field potentials, and with the reduced LdG predictions in 2D, for large $\lambda$ or equivalently, for sufficiently large $R$ and $U\gg U_{\text{c}}$. The consistency holds in equilibrium and non-equilibrium scenarios which is a noteworthy observation, although more conclusive and comprehensive studies are needed for non-equilibrium scenarios.

Our work demonstrates that the N-MPCD method is versatile and can work for different choices of the mean-field potentials, which is useful for computational scientists. Our results affirm the robustness of the continuum LdG approach (for this model problem) for large well-ordered and experimentally relevant systems, independently of the underlying mean-field model, which is not a priori obvious. It also provides a useful framework for defining limits of validity for continuum theories. For example, our methods can be generalised to compute $U_{\text{c}}$ and the nematic coherence length for arbitrary potentials, perhaps even for complicated molecular shapes outside the remit of standard rod-shaped uniaxial nematic molecules. As such, these methods can yield a systematic way of determining critical values, $R_{\text{c}}$ and $U_{\text{c}}$, such that the mean-field predictions agree with the corresponding coarse-grained continuum description for $R \geq R_{\text{c}}$ and $U\geq U_{\text{c}}$. This of course, lends itself to the question - what happens for $R < R_{\text{c}}$ and $U < U_{\text{c}}$? Do mean-field predictions differ substantially from continuum descriptions in this regime or are the mean-field predictions stochastic fluctuations of the continuum predictions for small and weakly-ordered systems? Can such small-size effects be exploited for applications? For example, can we have different predictions for defect-core structures according to mean-field models and if so, how do we choose the right mean-field model? Finally, we can also use our methods to design multiscale numerical methods, \textit{i.e.}, we could use N-MPCD methods near critical temperatures, defects and interfaces and couple them to deterministic LdG solvers in regular regions, using the mappings between LdG and N-MPCD parameters to define the coupling criteria. Hence, foundational studies, such as ours, are crucially needed to understand and quantify the multiphysics nature of NLC and soft matter phenomena.

\section{The Reduced Landau-de Gennes Model in 2D}
\label{sec:rLdG}
In this section, we review the reduced Landau-de Gennes model (rLdG model) for NLCs on 2D domains, $\Omega$.  In the Landau-de Gennes (LdG) theory, the macroscopic state of the NLC configuration is described by the LdG $\Qvec$-tensor order parameter: a $3\times 3$ symmetric matrix whose eigenvectors model the distinguished nematic directors and the corresponding eigenvalues measure the degree of orientational ordering about the corresponding eigenvectors/directors. The isotropic phase is modelled by $\Qvec=0$ and a nematic phase is labelled by $\Qvec \neq 0$. The physically observable configurations are modelled by local or global minimisers of an appropriately defined LdG free energy, typically a nonlinear and non-convex functional of $\Qvec$ and $\nabla \Qvec$, subject to the imposed boundary conditions.

Consider a domain of the form $\Omega \times [0,h]$, where $\Omega \in \mathbb{R}^2$ is a regular 2D domain and $h$ is much smaller than the cross-sectional dimensions of $\Omega$. In~\cite{golovatymontero2015}, the authors show that in the $h \to 0$ limit, and for certain choices of the boundary conditions such as tangent or planar degenerate boundary conditions on $z=0$ and $z=h$, the physically relevant critical points of the LdG free energy are $z$-invariant, have a fixed eigenvector in the $z$-direction or the vertical direction with an associated fixed eigenvalue. This automatically reduces the computational domain to $\Omega$ and there are only two degrees of freedom left in the LdG model, now referred to as the rLdG model in 2D.

In the rLdG model, the NLC phase is described by the rLdG $\matriz{Q}$-tensor: a symmetric, traceless $2 \times 2$ matrix, \textit{i.e.},
\[
\matriz{Q} = S\left\{ \unitvc{n}\otimes \unitvc{n} - \frac{\matriz{I}}{2} \right\}
\] where the nematic director, $\unitvc{n}(x,y) = \left(\cos\phi(x,y), \sin\phi(x,y) \right)$, is the eigenvector with the largest positive eigenvalue, $\matriz{I}$ is the $2\times 2$ identity matrix and $(x,y)\in\Omega$. The positive quantity, $S:\Omega \to \mathbb{R}$ (and $\unitvc{n}$ has eigenvalue $\frac{S}{2}$), is referred to as the \emph{scalar order parameter}. The director models the average preferred direction of alignment of the NLC molecules in the plane of $\Omega$, and the scalar order parameter measures the degree of directional ordering about $\unitvc{n}$. Thus, there are only two degrees of freedom in the rLdG model: one associated with the planar director, $\unitvc{n}$ and one associated with the scalar order parameter, $S$, and defects are associated with the nodal set of $S$~\cite{han_nonlinearity_2021}. In other words, a defect in the rLdG model is a point or a line wherein there is no nematic order in the plane of $\Omega$, defined by the nodal/zero set of $S$.

The physically observable configurations in experiments and applications are modelled by local or global minimisers of an appropriately re-scaled rLdG free energy in 2D~\cite{note_index_notation}:
\begin{equation}
\label{rLdGenergy}
I[\matriz{Q}] = \int_{\Omega} \frac{1}{2} |\nabla \matriz{Q}|^2 + \frac{\eta^2}{2}\left( \frac{A}{C}|\matriz{Q}|^2 + |\matriz{Q}|^4  \right)~dx_{1}\,dx_{2},
\end{equation}
where $\eta^2 \propto \frac{D^2 C}{L}$,  $D$ is a characteristic length scale of $\Omega$, $C$ is a positive material-dependent constant, $L$ is a positive material-dependent elastic constant and $A$ is the re-scaled temperature. In particular, $A <0$ in the nematic phase \cite{han_siam_j_appl_math_2020}. The first term is the one-constant elastic energy density that penalises spatial inhomogeneities in the system and the second polynomial term is a bulk potential that dictates the bulk nematic order as a function of the temperature coded in $A$. For example, the minimum of the bulk potential is the isotropic phase $\matriz{Q} = 0$, for $A>0$; and ordered nematic bulk minima appear for $A<0$. 
Note that the LdG free energy admits a cubic term in the bulk potential in 3D to capture the first order isotropic-nematic phase transition whereas the rLdG model predicts a second-order isotropic-nematic phase transition at $A=0$.

The functional \eqref{rLdGenergy} can admit multiple stable energy minimisers and non energy-minimising saddle points, as a function of $\eta$. In \cite{han_siam_j_appl_math_2020}, the authors study asymptotic profiles of minimisers of \eqref{rLdGenergy} on 2D regular polygons, subject to tangent boundary conditions, in two distinguished limits: the $\eta \to 0$ limit relevant for very small polygons e.g. nano-polygons for which $D \approx 10^{-9}\,\text{m}$, and the $\eta \to \infty$ limit valid for large polygons with $D$ of the order of microns or larger. The tangent boundary conditions require $\unitvc{n}$ to be tangent to the polygon edges and are translated to Dirichlet conditions for the rLdG order parameter on the polygon edges. 
In the $\eta \to 0$ limit, the authors prove that there is a unique critical point of \eqref{rLdGenergy}, labelled as the \emph{Ring} solution, on all polygons (except for the square and an equilateral triangle); the \emph{Ring} solution has a single isolated point defect (with $S=0$) at the polygon centre. 
On a square domain, the limiting profile is the Well Order Reconstruction Solution (WORS) with two isotropic lines along the square diagonals, in the $\eta \to 0$ limit. The corresponding director is constant in each square quadrant. On an equilateral triangle, the limiting profile has a $-1/2$-point defect at the centre, in the $\eta \to 0$ limit. In the $\eta \to \infty$ limit, the authors show that there are at least $[\frac{K}{2}]$ classes of rLdG minimisers on a $K$-polygon with $K$ edges, using variational and combinatorial arguments. For example, on a square with $K=4$, there are at least two classes of rLdG minimisers on square domains with tangent boundary conditions: the \emph{Diagonal} solution for which $\unitvc{n}$ is oriented along a square diagonal and the \emph{Rotated} solution for which $\unitvc{n}$ rotates by $\pi$-radians between a pair of parallel square edges.

In~\cite{yin_phys_rev_lett_2020}, the authors study the solution landscape of the rLdG free energy \eqref{rLdGenergy} on square domains, as a function of $\eta$ or equivalently as a function of the square edge length, with tangent boundary conditions on square edges. Solution landscapes are a connected network of critical points of \eqref{rLdGenergy}, that illustrate how unstable saddle points are connected to each other and how they connect stable critical points, and as such, contain crucial clues about the static and non-equilibrium transition pathways in multistable NLC systems. The authors label the critical points in terms of their \emph{Morse index}, \textit{i.e.}, the number of unstable directions of a critical point. The stable critical points or the energy minimisers have index-$0$, whereas transition states are typically index-$1$ saddle points. For example, the Morse index of the WORS increases with $\eta$ whereas the diagonal and rotated solutions have index-$0$, for sufficiently large $\eta$. Higher-index saddle points are typically distinguished by symmetric arrangements of interior $\pm 1/2$ and $\pm 1$-degree nematic defects. Transition states or index-$1$ saddle points are usually observed in switching processes and hence, have practical relevance, although switching can also be achieved efficiently by high-index saddle points~\cite{han_nonlinearity_2021}. In \cite{robinson_liq_cryst_2017}, the authors use deterministic (rLdG) models and molecular-level models (Gay-Berne models, Lebwohl-Lasher lattic models) to compute and compare admissible NLC configurations on square domains, with tangent boundary conditions. In the subsequent sections, we use the N-MPCD method and employ different mean-field models to compute admissible NLC configurations on square domains with tangent boundary conditions, including relaxation pathways, and compare with the deterministic rLdG predictions in existing literature. 

\section{\label{mean_field_potential_models_section}Mean-field Potential Models}

N-MPCD simulations are based on an orientation collision operator as described in the next section~\cite{shendruk_soft_matter_2015}; the constituent nematic particles or nematogens interact with a local macroscopic order parameter or a local mean-field and the collision operator dictates the reorientation of the nematogens post the mean-field interactions. In~\cite{hijar_soft_matter_2024}, the authors conduct N-MPCD simulations on regular polygons with the MS mean-field potential model. For 2D systems, the MS mean-field potential describes the local nematic state in terms of a mean-field order parameter, $\matriz{Q}$, that encodes information about the inter-particle interactions. The MS mean-field potential energy density is the energy associated with a nematogen that has orientation, $\hat{\mathbf{u}}$, in an environment with mean-field order parameter $\matriz{Q}$ according to~\cite{maier_z_naturforschung_1959}
\begin{equation}
V_{\text{MS}}\left( \hat{\mathbf{u}};\matriz{Q}\right) 
= -  \frac{1}{2} U k_{\text{B}} T 
\left( 
       u_{i} Q_{ij} u_{j} + 1
\right),
\label{model_001}
\end{equation}
where 
\begin{equation}
\Qvec = \left\{Q_{ij}\right\} 
= \left\langle 2 u_{i} u_{j} - \delta_{ij} \right\rangle ,
\label{model_002}
\end{equation}
is defined as an appropriately defined statistical average of $\uvec\otimes\uvec$. The parameter, $U$,
quantifies the strength of the inter-particle interactions and larger values of $U$ are associated with stronger nematic ordering. 

A critical modification of the MS theory, including  spatial gradients in $\matriz{Q}$, was proposed by Marrucci and Greco in~\cite{marrucci_mol_cryst_liq_cryst_1991}. This enables the study of phenomena such as director distortions, defects, and elasticity in NLCs. The MG mean-field potential energy density is given by:
\begin{equation}
V_{\text{MG}}\left( \hat{\mathbf{u}};\matriz{Q},\nabla^{2}\matriz{Q}\right) 
=  -\frac{U k_{\text{B}} T}{2} 
\left[ 
       u_{i} \left( Q_{ij} + \frac{l^{2}_{\text{MG}}}{24} \nabla_{k} \nabla_{k} Q_{ij}
       \right) u_{j} + 1
\right],
\label{model_004}
\end{equation}
where $l_{\text{MG}}$ denotes a characteristic length scale  for the molecular interaction; $\nabla_{k} = \partial/\partial x_{k}$ denotes the spatial derivative and $\matriz{Q}$ has been defined above. $V_{\text{MG}}\left(\hat{\mathbf{u}}\right)$ can account for spatial inhomogeneities and as such, might be better suited for simulations of confined or frustrated NLC systems. 

There are alternative mean-field potentials in the context of kinetic models for NLCs, such as the FPE for the orientational distribution function of molecules~\cite{ilg_phys_rev_e_1999,kroeger_j_chem_phys_2007,kroeger_j_nonnewton_fluid_mech_2008}. These efforts focus on systematically deriving hydrodynamic and mesoscopic models from microscopic principles, bridging the gap between molecular-level interactions and macroscopic phenomena. Further, Ilg, Karlin, and \"Ottinger have proposed a self-consistent mean-field potential for 2D nematics~\cite{ilg_phys_rev_e_1999}
\begin{equation}
V_{\text{IK\"O}}\left( \hat{\mathbf{u}};\matriz{Q}\right) 
= \frac{1}{2} U k_{\text{B}} T 
\frac{1 - u_{i} Q_{ij} u_{j}}
{\sqrt{2-Q_{ij} Q_{ij}}},
\label{model_003}
\end{equation}
which is expressed in terms of the same quantities appearing in Eq.~(\ref{model_001}). The minus sign of $u_{i}Q_{ij}u_{j}$ in Eq.~(\ref{model_003}) guarantees that $V_{\text{IK\"O}}$ possesses a minimum when $\unitvc{u}$ is parallel to the local average director (the eigenvector of $\Qvec$ with the positive eigenvalue - see \eqref{model_002}.) The potential $V_{\text{IK\"O}}\left( \hat{\mathbf{u}}\right)$ provides an approximate representation of the Onsager excluded-volume interaction and is particularly suitable for describing orientational order in suspensions of highly elongated, rigid, rod-like particles~\cite{ilg_phys_rev_e_1999,kroeger_j_nonnewton_fluid_mech_2008}. The corresponding relaxation equation for the order parameter, as derived from $V_{\text{IK\"O}}\left( \hat{\mathbf{u}}\right)$, differs from the MS counterpart~\cite{ilg_phys_rev_e_1999}. 

In what follows, we use the N-MPCD simulation method to simulate equilibrium NLC configurations and relaxation dynamics in square domains, with tangent boundary conditions, using the three representative mean-field potentials above. This enables us to capture  characteristics of a wide range of nematic-like materials, spanning from thermotropics to dilute or semi-dilute dispersions of rigid polymers. The N-MPCD simulation technique, described in detail in Section~\ref{simulations_section}, captures the dynamics of confined 2D nematics as an ensemble of interacting particles with time-dependent positions, velocities and orientations. The implementation leverages cell-based collision rules, boundary conditions, and probabilistic orientation updates to reproduce nematohydrodynamic phenomena, thereby offering a powerful approach to study the interplay between mean-field potentials and confinement effects for this prototype problem.

\section{Simulations \label{simulations_section}}

N-MPCD simulates the NLC system as an ensemble of $N$ point particles with mass $m$, position $\mathbf{r}_{\alpha}$, and velocity $\mathbf{v}_{\alpha}$ for $\alpha=1,2\dots, N$. These particles, hereafter referred to as \emph{nematogens}, carry an orientation unit vector $\hat{\mathbf{u}}_{\alpha}$. The vectors $\mathbf{r}_{\alpha}$, $\mathbf{v}_{\alpha}$, and $\hat{\mathbf{u}}_{\alpha}$ are time-dependent and restricted to the Cartesian plane $x_{1}x_{2}$. There are two main steps, known as the \emph{streaming} and \emph{collision} steps. Nematogens undergo ballistic motion over $n_{\text{stream}}$ time-steps of size $\delta t$, after which a collision event takes place that promotes exchanges of velocities and orientations. N-MPCD is designed to reproduce nematohydrodynamics after repetition of these streaming and collision steps~\cite{shendruk_soft_matter_2015}.

\begin{figure}
    \centering
    \includegraphics[width=0.95\linewidth]{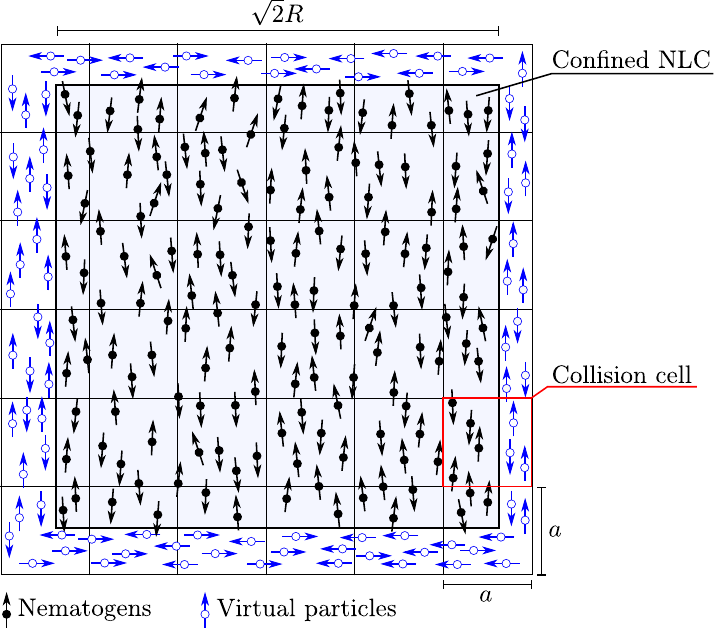}
    \caption{Schematic of the simulation model. Small black circles with an arrow represent N-MPCD particles/nematogens. They are confined in the light-blue area whereas the extended region contains virtual particles (small empty circles with an arrow). The virtual particles are used to enforce tangent boundary conditions and avoid collision in partially empty cells.}
    \label{figure_001}
\end{figure}

In our simulations, the computational NLC-filled region is a regular square domain with diagonal size $2R$. This region is illustrated as the light blue area in figure~\ref{figure_001}. 
The initial positions of the nematogens are randomly sampled; their initial velocities are randomly assigned from the Maxwell distribution at temperature $T$, and the center of mass velocity of the whole system is set to zero~\cite{frenkel_understanding_molecular_simulation_2002}. Initial orientations of the N-MPCD particles are assigned randomly by sampling initial $\phi_{\alpha}$ values from the uniform probability distribution in the range $(0,2\pi]$. To simulate confinement, bounce-back boundary conditions are applied for nematogens that collide with polygons sides during streaming. Also, such particles are reoriented to favour tangential anchoring conditions so that the nematogens are aligned parallel to the square edges.

At regular times separated by the interval $\Delta t$, the simulation space is subdivided into square cells of side $a$, hereafter referred to as \emph{collision cells}. The location of the grid is selected randomly from a uniform distribution in the range $[-a/2,a/2]$ along each Cartesian axis, to preserve Galilean invariance~\cite{ihle_phys_rev_e_2001}. The space outside the polygon if filled with virtual particles. These are auxiliary particles of mass $m$ uniformly distributed to have the same density as particles inside the polygon. Their velocities are taken from a Maxwell distribution at temperature $T$ with zero mean. Virtual particles also carry a unit vector representing their orientation. This vector is forced to be parallel to the polygon's side intersected by the line going from the center of the polygon to the particle. A schematic of the system at the collision step is shown in figure~\ref{figure_001}.

Each collision cell has a different number of particles, $N^{\text{c}}$, that includes nematogens and virtual particles. They contribute to cell-level properties as the center of mass, $\mathbf{r}^{\text{c}}$; center of mass velocity, $\mathbf{v}^{\text{c}}$, and moment of inertia, $\matriz{J}^{\text{c}}$, defined as usual in classical mechanics~\cite{symon_mechanics}. They also produce a local rLdG $\Qvec$-tensor
\begin{equation}
\matriz{Q}^{\text{c}} = \frac{1}{N^{\text{c}}} 
                        \sum_{\alpha \in \text{c}} 
                        \left( 2 \hat{\mathbf{u}}_{\alpha} \otimes \hat{\mathbf{u}}_{\alpha} -\matriz{I} \right),
\label{simulations_001}
\end{equation}
where 
$N^{\text{c}}$ the number of particles in the cell at the multi-particle collision instant. The local scalar order parameter, $S^{\text{c}}$, and director, $\unitvc{n}^{\text{c}}$, are the largest eigenvalue of $\matriz{Q}^{\text{c}}$ and the corresponding eigenvector respectively. In Eq.~(\ref{simulations_001}), summation extends over all particles located within cell $\text{c}$.

In the collision step, the nematogens in a collision cell interact through collective operators that update their positions and velocities. We assume that $\unitvc{u}_\alpha$ interacts with the local director through the prescribed mean field potential, $V$, in such a way that new orientations, $\unitvc{u}^{\prime}_{\alpha}$, are sampled from the corresponding canonical distribution 
\begin{equation}
    P\left(\unitvc{u}^{\prime}_{\alpha}\right) \propto \exp\left(-V\left(\unitvc{u}^{\prime}_{\alpha}\right)/\left(k_{B}T\right)\right), 
    \label{collision_step_003a}
\end{equation}
where $V$ is the mean-field potential. In the MS and IK\"O models, $V\left(\unitvc{u}^{\prime}_{\alpha}\right)$ is taken as the cell-level version of Eqs.~(\ref{model_001}) and (\ref{model_003}), \textit{i.e.}, $V\left(\unitvc{u}^{\prime}_{\alpha}\right)=V_{\text{MS}}\left(\unitvc{u}^{\prime}_{\alpha};\matriz{Q}^{\text{c}}\right)$ and $V\left(\unitvc{u}^{\prime}_{\alpha}\right)=V_{\text{IK\"O}}\left(\unitvc{u}^{\prime}_{\alpha};\matriz{Q}^{\text{c}}\right)$.

In practice, we use the relation between $\matriz{Q}$, the nematic director, $\unitvc{n}$, and the scalar order parameter, $S$,
\begin{equation}
\matriz{Q} = S\left(2\, \unitvc{n} \otimes \unitvc{n} -\matriz{I} \right),
\label{simulations_002a}
\end{equation}
in the local form $\matriz{Q}^{\text{c}} = S^{\text{c}}\left(2\, \unitvc{n}^{\text{c}} \otimes \unitvc{n}^{\text{c}} -\matriz{I} \right)$, where  $S^{\text{c}}$ and $\unitvc{n}^{\text{c}}$ have been defined above. The canonical probabilities can be written in terms of $\theta^{\prime}$, the angle between $\unitvc{u}^{\prime}_{\alpha}$ and $\unitvc{n}^{\text{c}}$. Straightforward algebra yields the unified form 
\begin{equation}
P\left(\theta^\prime \right) d\theta^{\prime} 
= C \exp\left(U_{\text{mf}}(S^{\text{c}}) \, S^{\text{c}} \cos^{2}\theta^{\prime}\right) d\theta^{\prime}, 
\label{simulations_005}
\end{equation}
where we have retained only the dependence on $\theta^{\prime}$ in the argument of the exponential, absorbed the remaining contributions in the normalizing constant $C$, and defined
\begin{equation}
U_{\text{mf}}(S) = \begin{cases}
    U, & \text{for the MS potential}; \\
    U/\sqrt{2\left(1-S^{2}\right)}, & \text{for the IK\"O potential} .
    \end{cases}
    \label{ldg_coefficients_004aa}
\end{equation}

Reorientation in the MG framework is handled in a similar fashion. Finite differences over the grid of collision cells are used to estimate spatial variations of the aforementioned fields. For instance, in the cell centered at $\left(x_{m},x_{n}\right)$, the second derivatives of $\matriz{Q}^{\text{c}}$ along $x_{1}$ are approximated by
\begin{equation}
\frac{\partial^{2}  \matriz{Q}^{\text{c}} }{\partial x_{1}^{2}} 
\simeq \frac{\matriz{Q}^{\text{c}}\left(x_{m} + a, x_{n}\right)-2 \matriz{Q}^{\text{c}}\left(x_{m},x_{n} \right)+\matriz{Q}^{\text{c}}\left(x_{m} - a,x_{n}\right)}{a^{2}}.
\label{simulations_002}
\end{equation}
Then the symmetric traceless tensor $\matriz{R}^{\text{c}}=\matriz{Q}^{\text{c}}+ l^{2}_{\text{MG}}\nabla^{2}\matriz{Q}^{\text{c}}/24$ is represented in the form, $\matriz{R}^{\text{c}} = \lambda_{+}^{\text{c}}\left( 2 \unitvc{m}^{\text{c}} \otimes \unitvc{m}^{\text{c}} -\matriz{I}\right)$, where $\lambda_{+}^\text{c}$ is its largest eigenvalue and $\unitvc{m}^{\text{c}}$ the corresponding eigenvector. Then the updated orientations, $\vartheta^{\prime}$, are defined by the angle between $\unitvc{u}^{\prime}_{\alpha}$ and $\unitvc{m}^{\text{c}}$,  which obey the probability distribution
\begin{equation}
P\left(\vartheta^\prime \right) d\vartheta^{\prime} 
= C \exp\left(U \lambda_{+}^{\text{c}} \cos^{2}\vartheta^{\prime}\right) d\vartheta^{\prime}. 
\label{simulations_007}
\end{equation}

The mean velocity in a collisions cell is given by
\begin{equation}
    \mathbf{v}^{\text{c}} = \frac{1}{N^{\text{c}}} 
                            \sum_{\alpha\in \text{c}}
                            \mathbf{v}_{\alpha}.
    \label{collision_step_002}
\end{equation}
The nematogens are further reoriented by flow according to
\begin{equation}
\frac{\unitvc{u}_{\alpha}^{\prime\prime}}{\Delta t} 
= \frac{\unitvc{u}_{\alpha}^{\prime}}{\Delta t} 
 + \chi_{\text{HI}} 
   \Bigl\{ \matriz{W}^{\text{c}} \cdot \unitvc{u}_{\alpha}^{\prime} 
         + \lambda_{\text{t}} 
            \left[ \matriz{D}^{\text{c}} \cdot \unitvc{u}_{\alpha}^{\prime} 
                  +\matriz{D}^{\text{c}} : \left( \unitvc{u}_{\alpha}^{\prime} 
                  \otimes \unitvc{u}_{\alpha}^{\prime} \right) \unitvc{u}_{\alpha}^{\prime}
            \right]
   \Bigr\}, 
\label{simulations_008}
\end{equation}
where $\Delta t$ is the time interval separating two collision events; $\matriz{W}^{\text{c}}$ and $\matriz{D}^{\text{c}}$ are the skew-symmetric and symmetric components of the velocity gradient in the cell, $\left(\boldsymbol{\nabla}\mathbf{v}\right)^{\text{c}}$ respectively; and $:$ indicates the double inner product. For the latter, spatial derivatives are approximated using finite differences over the grid of collision cells as, \textit{e.g.},
\begin{equation}
\frac{\partial\, \mathbf{v}^{\text{c}}}{\partial x_{1}}\left(x_{m},x_{n}\right) 
\simeq \frac{ \mathbf{v}^{\text{c}}\left(x_{m} + a,x_{n}\right) - \mathbf{v}^{\text{c}}\left(x_{m} - a, x_{n}\right)}{2 a}.
\label{simulations_009}
\end{equation} We note that Eq.~(\ref{simulations_008}) is a discrete version of the Jeffery's model for reorientation of slender rods under flow~\cite{jeffery_proc_r_soc_lond_a_1922}, where $\lambda_{\text{t}}$ is the tumbling parameter, and $\chi_{\text{HI}}$ is a parameter that adjusts the relaxation time of reorientation by flow with respect to $\Delta t$. Reorientation by flow is turned off by imposing $\chi_{\text{Hi}}=0$.   
An additional collision operator assigns new velocities to nematogens, $\mathbf{v}_{\alpha}^{\prime}$, according to the Andersen rule, that guarantees linear and angular momentum conservation,
\begin{equation}
\mathbf{v}_{\alpha}^{\prime} = \mathbf{v}^{\text{c}} + \boldsymbol{\xi}_{\alpha} -\boldsymbol{\xi}^{\text{c}} + \left( \matriz{J}^{\text{c}}\right)^{-1} \cdot \Delta \mathbf{L}^{\text{c}} \times \left( \mathbf{r}_{\alpha} - \mathbf{r}^{\text{c}} \right),
\label{simulations_010}
\end{equation}
where $\boldsymbol{\xi}_{\alpha}$ is a random velocity sampled from the 2D Maxwell velocity distribution with zero mean and standard deviation $\sqrt{k_{\text{B}}T/m}$; $\boldsymbol{\xi}^{\text{c}}=\sum_{\alpha \in \text{c}} \boldsymbol{\xi}_{\alpha}/N^{\text{c}}$; and $\Delta \mathbf{L}^{\text{c}}$ is the angular momentum generated by velocity changes, $\Delta \mathbf{L}^{\text{c}}_{\text{vel}}=m\sum_{i \in \text{c}} \left(\mathbf{r}_{\alpha}-\mathbf{r}^{\text{c}}\right)\times\left(\mathbf{v}_{\alpha}-\mathbf{v}^{\text{c}}\right)$, plus a contribution due to dissipative orientation torques, $\Delta \mathbf{L}^{\text{c}}_{\text{ori}} = \sum_{\alpha\in \text{c}} \boldsymbol{\Gamma}_{\alpha}^{\text{diss}}\Delta t$. To model the latter, it is considered that reorientation is overdamped. Thus, torques on nematogens due to the collision operator, $\boldsymbol{\Gamma}_{\alpha}^{\text{col}}$, reorientation by flow, $\boldsymbol{\Gamma}_{\alpha}^{\text{HI}}$, and dissipation, $\boldsymbol{\Gamma}_{\alpha}^{\text{diss}}$, must be balanced
\begin{equation}
\boldsymbol{\Gamma}_{\alpha}^{\text{col}}+\boldsymbol{\Gamma}_{\alpha}^{\text{HI}}+\boldsymbol{\Gamma}_{\alpha}^{\text{diss}}= 0 .
\label{simulations_011}
\end{equation}

By introducing another parameter representing rotational friction, $\gamma_{\text{R}}$, one has $\boldsymbol{\Gamma}_{\alpha}^{\text{diss}} = -\left(\boldsymbol{\Gamma}_{\alpha}^{\text{col}}+\boldsymbol{\Gamma}_{\alpha}^{\text{HI}} \right)= -\gamma_{\text{R}} \unitvc{u}_{\alpha} \times \unitvc{u}_{\alpha}^{\prime\prime} / \Delta t$. 

Nematogens receive an impulse from reorientation by including $\Delta \mathbf{L}^{\text{c}}_{\text{ori}}$ in Eq.~(\ref{simulations_010}). The total impulse in the cell created by this mechanism is a reorientation-induced flow and represents the way in which N-MPCD accounts for backflow.  

Finally, the tangent boundary conditions are implemented by means of three conditions, previously used to generate anchoring conditions on colloidal surfaces~\cite{hijar_phys_rev_e_2020,reyes_physica_a_2020,duarte_physica_a_2023}. The orientations of virtual particles are prescribed to be parallel to the square edges (see Fig.~\ref{figure_001}). N-MPCD particles that collide with the perimeter of the polygon are reoriented in the direction of the edge. 
Virtual particles can participate in the orientation collision operator and favour $\unitvc{n}^{\text{c}}$ to be tangent to the square edges, or near the square edges.

\section{Applicability limits and relation with a macroscopic model \label{applicability_limits_section}}
It is essential to identify the values of the interaction parameter $U$ in \eqref{model_001}, \eqref{model_003} and \eqref{model_004}, for which the system exhibits nematic ordering as opposed to isotropic or disordered patterns, in order to define the range of applicability of our N-MPCD simulations and to interpret them in the framework of continuum theory. To achieve this, we model N-MPCD systems as microstructured fluids and the collective dynamics of nematogens can be coarse-grained to yield macroscopic evolution equations, analogous to those derived from the continuum rLdG theory.

We characterize the stochastic nature of the N-MPCD framework using the probability distribution function $g\left(\unitvc{u},t\right)$, which models the likelihood of finding a nematogen  oriented along direction $\unitvc{u}$, at time $t$. The time evolution of this distribution can be described within a Fokker–Planck (FP) formalism that incorporates the effects of stochastic collisions, mean-field orientational interactions, and rotational diffusion. Accordingly, we propose that the dynamics of $g$ is governed by the following kinetic equation 
\begin{equation}
    \frac{\partial g}{\partial t}
    = D_{\text{r}} \, \mathbf{R} \cdot \left[\mathbf{R} g 
    + \frac{g}{k_{B} T} \mathbf{R} V_{\text{mf}}\right] 
    ,
    \label{relation_macro_001}
\end{equation}
which can be coarse-grained for the derivation of macroscopic continuum descriptions.

The right-hand side of Eq.(\ref{relation_macro_001}) represents the balance between alignment interactions, encoded in the mean-field potential $V_{\text{mf}}$, and the disordering effects of rotational diffusion, which is assumed to be isotropic with diffusion coefficient $D_{\text{r}}$. The operator $\mathbf{R} = \unitvc{u}\times\partial/\partial\unitvc{u}$ denotes the rotational derivative~\cite{kroeger_j_chem_phys_2007}. In our 2D system, this operator simplifies due to the constraints $u_{3} = 0$ and $\partial/\partial\, u_{3} = 0$.

In the following subsection, we use the FPE~(\ref{relation_macro_001}) to derive evolution equations for the key macroscopic nematic fields: the order parameter tensor field and the scalar order parameter. 

\subsection{Order parameter evolution equation}
\label{order_parameter_evolution_section}

To bridge the mean-field models with the continuum rLdG description, we construct a hierarchy of equations for the  moments of the orientational distribution function $g$, starting from Eq.~(\ref{relation_macro_001})~\cite{hess_z_naturforsh_a_1976,doi_1986,kroger_j_chem_phys_1995,ilg_phys_rev_e_1999,kroeger_j_chem_phys_2007,kroeger_j_nonnewton_fluid_mech_2008}. The mean-field nematic order parameter is defined to be the second moment of $g$ as given below:
\begin{equation}
Q_{ij} =  \int_{S^1} d\unitvc{u}\, \left(2 u_{i}u_{j} -\delta_{ij}\right)\, g\left(\unitvc{u}\right),
\label{ldg_coefficients_003}
\end{equation}
where $\delta_{ij}$ is the Kronecker delta symbol and $S^1$ is the unit circle in 2D.

The evolution equation for $Q_{ij}$ is obtained by multiplying Eq.~(\ref{relation_macro_001}) by $2u_{i}u_{j} - \delta_{ij}$, and integrating over $S^1$. For consistency with approximations for spatially homogeneous systems, we neglect spatial gradients and obtain
\begin{eqnarray}
\frac{\partial Q_{ij}}{\partial t} & = &
- 4 D_{\text{r}} \left[1-\frac{U_{\text{mf}}\left(S\right)}{2}\right] Q_{ij} 
+ 2 D_{\text{r}} U_{\text{mf}}\left(S\right) Q_{ik} Q_{kj} \nonumber \\
&  & - 4 D_{\text{r}} U_{\text{mf}}\left(S\right) 
     \langle u_{i} u_{j} u_{k} u_{l} \rangle Q_{kl} ,
\label{ldg_coefficients_004}
\end{eqnarray}
where $U_{\text{mf}}(S)$ is identical for the MS and MG cases. Equation~(\ref{ldg_coefficients_004}) couples $Q_{ij}$ with the fourth-order average $\langle u_{i} u_{j} u_{k} u_{l} \rangle$, and is the first equation of the aforementioned hierarchy. 
Such hierarchical schemes are ubiquitous in the analysis of structured fluids~\cite{hess_z_naturforsh_a_1976,hinch_j_fluid_mech_1976,kroger_j_chem_phys_1995,ilg_phys_rev_e_1999,kroeger_j_chem_phys_2007,kroeger_j_nonnewton_fluid_mech_2008}. Since the hierarchy is infinite (the fourth-order moment couples to the sixth-order moment, and so on), practical solutions typically rely on closure relations that approximate higher-order moments in terms of lower moments of $g$. We use the following closure relation for the fourth-order moment as derived in appendix~\ref{appendix_001}:
\begin{eqnarray}
\langle u_{i} u_{j} u_{k} u_{l} \rangle & = & 
\alpha 
\left\{ \langle \unitvc{u} \otimes \unitvc{u} \rangle \otimes
        \langle \unitvc{u} \otimes \unitvc{u} \rangle \right\}^{\text{sym}}_{ijkl} \nonumber \\
&  & -2\beta 
\left\{ \langle \unitvc{u} \otimes \unitvc{u} \otimes \matriz{I} \rangle \right\}^{\text{sym}}_{ijkl} \nonumber \\
&  & -2\gamma 
\left\{ \matriz{I} \otimes \matriz{I} \right\}^{\text{sym}}_{ijkl}.
    \label{ldg_coefficients_006}
\end{eqnarray}

Here, the coefficients $\alpha$, $\beta$, and $\gamma$ are defined by
\begin{equation}
\alpha = \frac{S_{4}}{S^{2}}, 
-2\beta= 1-\frac{S_{4}}{S^{2}}, 
\text{ and} -2\gamma =  \frac{S_{4}}{4 S^{2}} - \frac{S_{4}}{8} -\frac{1}{8};
\label{ldg_coefficients_007}
\end{equation}
in terms of $S$ and $S_{4} = \langle \cos\left(4 \theta\right)\rangle$---the fourth-order scalar order parameter characterizing the 2D nematic phase, where $\theta$ the angle between $\unitvc{u}$ (the orientation of a nematogen) and $\unitvc{n}$ (the local nematic director or leading eigenvector of the mean-field order parameter defined in \eqref{ldg_coefficients_003}).

The explicit form of tensors in Eq.~(\ref{ldg_coefficients_006}) can be obtained from Eq.~(\ref{details_appendix_005}) and reads as
\begin{eqnarray}
\left\{ \langle \unitvc{u} \otimes \unitvc{u} \rangle \otimes 
        \langle \unitvc{u} \otimes \unitvc{u} \rangle \right\}^{\text{sym}}_{ijkl}
& = & \frac{1}{3} 
\left[ \langle u_{i} u_{j} \rangle \langle u_{k} u_{l} \rangle
    +  \langle u_{i} u_{k} \rangle \langle u_{j} u_{l} \rangle \right. \nonumber \\
&   & \left.    +  \langle u_{i} u_{l} \rangle \langle u_{j} u_{k} \rangle
\right] ,
\label{ldg_coefficients_008}    
\end{eqnarray}
\begin{eqnarray}
\left\{ \langle \unitvc{u} \otimes \unitvc{u} \otimes \matriz{I} \rangle \right\}^{\text{sym}}_{ijkl}
& = &  \frac{1}{6}
\left[ \langle u_{i} u_{j} \rangle \delta_{kl}
    +  \langle u_{i} u_{k} \rangle \delta_{jl}  \right. \nonumber \\ 
&   &  \left.     +  \langle u_{i} u_{l} \rangle \delta_{jk}
    +  \langle u_{j} u_{k} \rangle \delta_{il}  \right. \nonumber \\ 
&   &  \left.    +  \langle u_{j} u_{l} \rangle \delta_{ik} 
    +  \langle u_{k} u_{l} \rangle \delta_{ij} 
\right],
\label{ldg_coefficients_009}    
\end{eqnarray}
and
\begin{equation}
\left\{ \matriz{I} \otimes \matriz{I} \right\}^{\text{sym}}_{ijkl} = \frac{1}{3} 
\left[ \delta_{ij} \delta_{kl} 
    +  \delta_{ik} \delta_{jl} 
    +  \delta_{il} \delta_{jk} 
\right] .
\label{ldg_coefficients_010} 
\end{equation}

Eqs.~(\ref{ldg_coefficients_006}) to (\ref{ldg_coefficients_010}) can be viewed as the 2D analogue of the consistent closure proposed by Kr\"oger, Ammar, and Chinesta for 3D uniaxial nematics~\cite{kroeger_j_nonnewton_fluid_mech_2008}. This closure satisfies the important consistency condition
\begin{equation}
\text{Tr}\left( \langle \unitvc{u} \otimes \unitvc{u} \otimes 
                        \unitvc{u} \otimes \unitvc{u} \rangle\right)
= \langle \unitvc{u} \otimes \unitvc{u} \rangle,
\label{ldg_coefficients_011}
\end{equation}
ensuring that contraction over any pair of indices recovers the second order moment of $g$.

The proposed closure model must be complemented by an approximate functional relation $S_{4}=S_{4}(S)$. An specific relation of this type will be given in section~\ref{scalar_closure_relation_section} and this function must be properly bounded within the admissible range of $S$. Using Eq.~(\ref{ldg_coefficients_003}) and Eqs.~(\ref{ldg_coefficients_006}) to (\ref{ldg_coefficients_010}), Eq.~(\ref{ldg_coefficients_004}) can be cast in the following closed form -
\begin{eqnarray}
\frac{\partial Q_{ij}}{\partial t} 
& = & -4 D_{\text{r}} \left( 1-\frac{U_{\text{mf}}(S)}{4} -\frac{U_{\text{mf}}(S) S_{4}(S)}{12} \right) Q_{ij} \nonumber \\
&   &
-\frac{D_{\text{r}} U_{\text{mf}}(S) S_{4}(S)}{3 S^{2}} Q_{kl}\left( Q_{lk} Q_{ij} + 2 Q_{ik}Q_{lj} \right).
    \label{ldg_coefficients_012}
\end{eqnarray}

The corresponding evolution equation for the scalar order parameter can be obtained by multiplying Eq.~(\ref{ldg_coefficients_012}) with $n_{i}n_{j}$, \textit{i.e.}, $Q_{ij}n_i n_j = S$. This yields  
\begin{equation}
\frac{\partial S}{\partial t} = -4D_{\text{r}} \left(1-\frac{U_{\text{mf}}(S)}{4}\right) S 
- D_{\text{r}} S \, U_{\text{mf}}(S) \, S_{4}(S),
\label{ldg_coefficients_017}
\end{equation}
which can be used to model the onset of nematic ordering.

Consider the stationary condition $\partial S/\partial t =0$. Then we either have $S=0$ or  
\begin{equation}
1 - \frac{U_{\text{mf}}(S)}{4} \left[ 1 - S_{4}(S)\right] = 0.
\label{ldg_coefficients_018}
\end{equation}
In 2D, there is a second order isotropic-nematic phase transition so that ordered nematic solution branches of \eqref{ldg_coefficients_018} (with positive $S$) emerge continuously from the $S=0$ branch. We assume a general closure relation of the form:
\begin{equation}
\lim_{S\rightarrow 0+} S_{4}(S) = 0 ,
\label{ldg_coefficients_019}
\end{equation}
Then Eq.~(\ref{ldg_coefficients_018}) admits non-zero solutions, $S$, for $U_{\text{c}}>4$ with both the MS and MG mean-field potentials, and for $U_{\text{c}} > 4 \sqrt{2}\simeq 5.66$ with the IK\"O approximation. We interpret these estimates of $U_{\text{c}}$ as the critical interaction strength needed for the onset of nematic ordering in confined systems, in the subsequent sections.

\subsection{Scalar closure relation, model validation, and asymptotic limits}
\label{scalar_closure_relation_section}

To complete the theoretical model, we now propose a generic closure relation that expresses the fourth-order scalar order parameter, $S_{4}$, as a function of the second-order scalar order parameter $S$. In particular, this closure relation is independent of the FPE~\eqref{relation_macro_001} or the probability distribution function, $g$ in \eqref{relation_macro_001}. In 2D NLCs, the scalar order parameters are typically defined using trigonometric functions of the angular degree of freedom $\theta \in (0,2\pi]$, where the nematogen orientation $\uvec = \left(\cos\theta, \sin \theta \right)$~\cite{frenkel_statistical_mechanics_lcs_1991}. Therefore, the $m$-order scalar order parameter for a 2D NLC (where $m$ is even): 
\begin{equation}
S_{m} = \langle \cos\left(m\theta\right)\rangle
    \label{scalar_order_params_001}
\end{equation}
for $m=2,4,\dots$, and the average is taken with respect to a suitably defined probability distribution function as defined below. This choice is motivated by the Fourier expansion of the distribution function that respects the headless symmetry of the nematogens~\cite{frenkel_statistical_mechanics_lcs_1991}. With $S = S_{2}$, we have $S = \langle \cos \left(2\theta\right)\rangle$ and, as previously stated, $S_{4} = \langle \cos \left(4\theta\right)\rangle$.

We work with the 2D analogue of the so-called uniaxial Bingham-type distribution function for the nematogen orientations ~\cite{kroeger_j_nonnewton_fluid_mech_2008},
\begin{equation}
    P\left(\theta\right)\propto\exp\left(A (U-U_{\text{c}}) \cos(2\theta)\right),
\end{equation} 
where $A$ is a constant. 
Thus, $S$ and $S_{4}$ can be written as
\begin{equation}
S(U) = \frac{1}{Z(U)} \int_{0}^{2\pi} d\theta\, \text{e}^{A (U-U_{\text{c}})\, \cos{2\theta}} \cos{2\theta},
    \label{scalar_order_001}
\end{equation}
and
\begin{equation}
S_{4}(U) = \frac{1}{Z(U)} \int_{0}^{2\pi} d\theta\, \text{e}^{A (U-U_{\text{c}})\, \cos{2\theta}} \cos{4\theta},
    \label{scalar_order_002}
\end{equation}
where the normalization factor is given by
\begin{equation}
Z(U) =  \int_{0}^{2\pi} d\theta\, \text{e}^{A (U-U_{\text{c}})\, \cos{2\theta}}.
    \label{scalar_order_003}
\end{equation}

These integrals are evaluated in terms of modified Bessel functions of the first kind as~\cite{abramowitz_1965_handbook}
\begin{equation}
S(U) = \frac{I_1(A(U-U_{\text{c}}))}{I_0(A(U-U_{\text{c}}))},
\label{scalar_order_003a}
\end{equation}
\begin{equation}
S_4(U) = \frac{I_2(A(U-U_{\text{c}}))}{I_0(A(U-U_{\text{c}}))}.
\label{scalar_order_003b}
\end{equation}

We expand these expressions in a Taylor series around the transition point $U=U_{\text{c}}$ (as identified in the preceding section for the MS, MG and IK\"O potentials), and retain only the leading-order contributions to obtain the approximate relation $S_{4}\simeq S^{2}/2$. To improve this closure and ensure the correct asymptotic behavior in the ordered limit $S \to 1$, we incorporate an additional term proportional to $S^{4}$ and impose the natural condition
\begin{equation}
\lim_{S\rightarrow 1} S_{4}(S) = 1.    
\label{scalar_order_004}
\end{equation}
This yields the improved closure expression
\begin{equation}
S_{4}(S) = \frac{S^{2} + S^{4}}{2}
    \label{scalar_order_005}
\end{equation} such that
$\lim_{S\to 0} S_4(S) = 0$ and $\lim_{S\to 1} S_4(S) = 1$.
This closure relation for $S_4$ provides a smooth interpolation between the disordered and highly ordered regimes and will be used in the remainder of this work to close the evolution equations.

\begin{figure}
    \centering
    \includegraphics[width=0.95\linewidth]{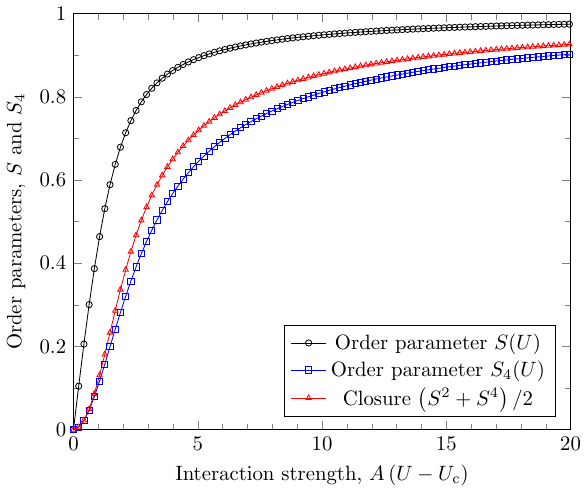}
    \caption{Comparison between the second and fourth scalar order parameters, computed from Eqs.(\ref{scalar_order_003a}) and (\ref{scalar_order_003b}), and the closure approximation from Eq.~(\ref{scalar_order_005}). The closure expression accurately captures the behavior of $S_{4}(U)$ near small values of $U-U_{\text{c}}$, while exhibiting a slight overestimation for large values of $U$.}
    \label{figure_002}
\end{figure}

In Fig.~\ref{figure_002}, we compare the exact analytic expressions for the scalar order parameters $S(U)$ and $S_{4}(U)$, as given by Eqs.~(\ref{scalar_order_003a}) and(\ref{scalar_order_003b}), with the closure relation in Eq.~(\ref{scalar_order_005}). The figure shows that the closure expression accurately captures the behavior of $S_{4}(U)$ near the critical interaction strength associated with the onset of nematic ordering, \textit{i.e.}, for $U\simeq U_{\text{c}}$. At higher values of $U$, that correspond to stronger ordering, the closure slightly overestimates $S_{4}$ and increases more rapidly toward the upper bound $S_{4}=1$.

To assess the validity of the closure scheme developed in Sections~\ref{order_parameter_evolution_section} and~\ref{scalar_closure_relation_section}, we now compare its predictions for $S(U)$ with those obtained from direct N-MPCD simulations of the MS, MG and IK\"O potentials. To this end, we substitute the closure relation given in Eq.(\ref{scalar_order_005}) into the equilibrium condition Eq.(\ref{ldg_coefficients_018}) and solve for $S(U)$. For the MS and MG schemes, we obtain the solution
\begin{equation}
S(U) = \sqrt{\frac{-1+\sqrt{1+8\left(1-4/U\right)}}{2}},
\label{scalar_order_006}
\end{equation}
whereas for the IK\"O potential, the solution is given by \begin{equation}
S(U) = 2\cos\left(\frac{\theta}{3}\right) - 1,
\label{scalar_order_007}
\end{equation}
with
\begin{equation}
\theta = \arctan
\left( \frac{ 2\sqrt{ \left(U_{\text{c}}/U\right)^{2}
                     -\left( U_{\text{c}}/U\right)^{4}}}{1-2\, \left(U_{\text{c}}/U\right)^{2}} 
\right) .
\label{scalar_order_008}
\end{equation}

N-MPCD simulations are carried out using units defined by $m=1$, $k_{\text{B}} T = 1$, and $a = 1$, corresponding to units of mass, energy, and length respectively. Time is measured in units of $u_{\text{t}} = a\sqrt{m/(k_{\text{B}}T)}$. Equilibrium values of the order parameter are computed in N-MPCD systems of size $l^{2}=50^{2} a^{2}$ with periodic boundary conditions. The following simulation parameters are used: average number of particles per cell $\bar{N}^{\text{c}} = 20$; advance and collision step-size, $\delta t = \Delta t = 0.1\,u_{\text{t}}^{-1}$; no coupling to hydrodynamic flow, $\chi_{\text{HI}} = 0$; and rotational friction coefficient, $\gamma_{\text{R}} = 0.01~u_{\text{t}}^{-1}$. Averages are taken from a time-series encompassing $\qty{2500}{}$ collision steps after thermalization.

\begin{figure}
    \centering
    \includegraphics[width=0.95\linewidth]{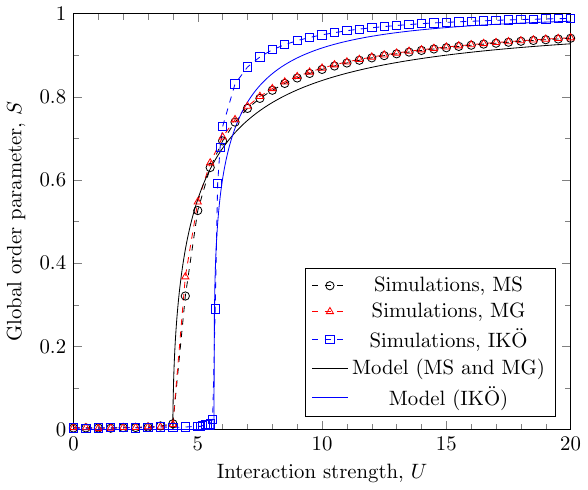}
    \caption{The emergence of nematic order near $U=U_c$ with different mean-field potential models, \textit{i.e.}, Maier–Saupe (MS), Marrucci–Greco (MG), and Ilg–Karlin–Öttinger (IKÖ) mean-field potentials. The scalar order parameter, $S(U)$, is obtained from N-MPCD simulations (symbols) and analytical predictions based on \eqref{scalar_order_006} and \eqref{scalar_order_007}-\eqref{scalar_order_008} (continuous curves). The agreement across all models confirms the validity of the closure scheme and its ability to reproduce phase transitions and ordered profiles.}
    \label{figure_003}
\end{figure}

In Figure~\ref{figure_003}, we compare the equilibrium values of the scalar order parameter $S(U)$ obtained from N-MPCD simulations with the analytical predictions derived from the FP model. These predictions are based on the closure relations, Eqs.(\ref{ldg_coefficients_006}) to (\ref{ldg_coefficients_010}), and (\ref{scalar_order_005}). Together, these yield explicit analytical expressions for $S(U)$ under different mean-field potentials: Eq.~(\ref{scalar_order_006}) for the MS and MG potentials, and Eqs.~(\ref{scalar_order_007})–(\ref{scalar_order_008}) for the IK\"O potential.

Figure~\ref{figure_003} demonstrates good agreement between theory and simulation across a broad range of interaction strengths. The small deviations observed between the simulation results and the analytical predictions can be attributed to the approximations introduced by the closure scheme.  
In particular, the scalar closure relation, Eq.~(\ref{scalar_order_005}), though asymptotically consistent, is an empirical fit and does not capture the full richness of the orientation distribution function beyond the leading angular harmonics. As such, the small discrepancies are consistent with the level of approximation.
More importantly, the model accurately captures the location of the critical interaction strength, $U_{\text{c}}$ consistent with those derived in Section\ref{order_parameter_evolution_section} and the nonlinear growth of the $S(U)$ above the critical point, $U>U_{\text{c}}$. 

These numerical tests support the main objective of this section: to establish a lower bound for the interaction strength $U$ at which nematic ordering emerges in N-MPCD simulations. Having identified $U_{\text{c}}$ as the critical threshold ($U_{\text{c}}=4$ for MS and MG potentials, and $U_{\text{c}}=4\sqrt{2}$ for the IK\"O potential), we carry out simulations for $U\geq U_{\text{c}}$ in subsequent sections to study confinement effects in a mean-field and N-MPCD framework.

\subsection{Normalized length scale}
\label{normalized_length_scale_section}

In Ref.~\cite{robinson_liq_cryst_2017}, confined NLC systems are characterized using the dimensionless effective normalized length scale
\begin{equation}
\lambda = \frac{l}{\xi_{\text{N}}} ,
    \label{normalized_length_001}
\end{equation}
where $l$ is the polygon/square edge length, $l= \sqrt{2} R$; and  $\xi_{\text{N}}$ is the nematic coherence length. The nematic coherence length quantifies the competition between elastic distortions and bulk ordering in 3D NLCs, and is given by
\begin{equation}
\xi_{\text{N}} = \sqrt{\frac{L}{A_{\text{NI}}}} ,
    \label{coherence_length_001}
\end{equation}
where $L$ is an elastic constant and $A_{\text{NI}} = \beta_{F}^{2}/12\gamma_{\text{F}}$ is a characteristic bulk coefficient constructed from the continuum LdG parameters  $\beta_{\text{F}}$ and $\gamma_{\text{F}}$, to be specifically defined below.
This coherence length serves as a material-dependent reference scale, often associated with defect core sizes and spatial variations in the order parameter near phase transitions~\cite{musevic_liquid_crystal_colloids_2017,trebin_liq_cryst_1998,trebin_chapter_2001,kleman_soft_matter_physics_2003}.

Thus, to quantitatively compare our N-MPCD simulation results for NLCs in square domains with tangent boundary conditions, with the deterministic continuum LdG numerical results in ~\cite{robinson_liq_cryst_2017,canevari_siam_2017,han_siam_2020}, it is essential to estimate the nematic coherence length $\xi_{\text{N}}$ as a function of the N-MPCD simulation parameters. The nematic coherence length can only be defined in a 3D context and hence, the methodology in Sections~\ref{order_parameter_evolution_section} and~\ref{scalar_closure_relation_section} needs to be extended to 3D. This generalization also allows us to derive coarse-grained dynamical equations that explicitly incorporate the interaction strength $U$. These coarse-grained equations can then be directly compared with the continuum LdG equations, that involve phenomenological LdG bulk and elastic coefficients. We can then match coefficients between the coarse-grained equations and the LdG equations, providing a consistent and predictive link between the simulation regime and theoretical expectations based on the normalized length scale, $\xi_{\text N}$. 
For concreteness, we only consider the MS mean-field potential and assume that the order of magnitude of $\xi_{\text{N}}$ is similar for the systems based on the MG and IK\"O interactions.

Several research groups have previously addressed the challenge of deriving continuum-level descriptions of confined 3D nematic systems from the underlying FPE, particularly through moment hierarchies involving orientational tensors~\cite{doi_1986,kroger_j_chem_phys_1995,chaubal_j_rheol_1998,feng_j_rheol_1998,kroeger_j_chem_phys_2007,kroeger_j_nonnewton_fluid_mech_2008,hijar_j_chem_phys_2012}. In these approaches, the structure and accuracy of the resulting macroscopic equations strongly depend on the closure scheme employed to truncate the hierarchy~\cite{chaubal_j_rheol_1995,kroeger_j_nonnewton_fluid_mech_2008}. Among the various strategies proposed, we adopt the theoretical framework developed by Kr\"oger and collaborators~\cite{kroeger_j_chem_phys_2007,kroeger_j_nonnewton_fluid_mech_2008}, which provides a consistent approximation for the fourth-order orientational tensor under the assumption of uniaxial symmetry. 

Neglecting spatial inhomogeneities, the model introduced in Refs.~\cite{kroeger_j_chem_phys_2007,kroeger_j_nonnewton_fluid_mech_2008} yields the following evolution equation for the scalar order parameter $S$ in 3D~\cite{hijar_j_chem_phys_2012}
\begin{equation}
\frac{\partial S}{\partial t} 
= -6 D_{\text{r}} 
\left\{ \left(1-\frac{U}{5} \right) S 
       -\frac{U}{7} S^{2}
       +\frac{12}{35} U S^{2}\left[1-\left(1-S\right)^{\nu}\right]
\right\},
\label{normalized_scale_001}
\end{equation}
where $\nu$ is a positive constant arising from the closure function $S_{4}(S) = S\left[1-\left(1-S\right)^{\nu}\right]$. The latter is obtained similarly to Eq.~(\ref{scalar_order_005}), by comparing the curves of  $S$ and $S_{4}$ resulting from a uniaxial 3D Bingham-type distribution~\cite{kroeger_j_nonnewton_fluid_mech_2008}.

On the other hand, the evolution of the LdG order parameter, $\matriz{Q}$, in the LdG framework is dictated by a gradient flow equation associated with the LdG bulk energy
\[ F_{b}(\Qvec) = \alpha_F \frac{Q_{ij}Q_{ij}}{2} - \beta_F Q_{ij}Q_{jk}Q_{ki} +  \gamma_F \left(Q_{ij}Q_{ij}\right)^2
\]
where $\alpha_F \to 0$ labels the onset of nematic ordering; see ~\cite{qian_phys_rev_e_1998,mandal_phys_rev_e_2019}. Note that $F_b$ is the simplest fourth-order polynomial that can capture a first-order isotropic-nematic phase transition in 3D; $\alpha_F$, $\beta_F$ and $\gamma_F$ are phenomenological material-dependent and temperature-dependent constants. The corresponding evolution equation for $\Qvec$ is given by:
\begin{equation}
\gamma_{1} \frac{\partial Q_{ij}}{\partial t} = -\alpha_{F} Q_{ij} +\beta_{F} \left(3 Q_{ik} Q_{kj} -Q_{kl} Q_{lk} \delta_{ij}\right) - 4\gamma_{F} Q_{kl}Q_{lk} Q_{ij}, 
\label{ldg_coefficients_002}
\end{equation}
where $\gamma_{1}$ is a rotational viscous coefficient. 
Assuming uniaxial symmetry with $Q_{ij}=S\left(3 n_{i}n_{j}-\delta_{ij}\right)/2$, and projecting onto $n_{i}n_{j}$, Eq.~(\ref{ldg_coefficients_002}) simplifies to
\begin{equation}
\gamma_{1}\frac{\partial S}{\partial t} 
= -\alpha_{\text{F}} S +\frac{3}{2} \beta_{\text{F}} S^{2} -6 \gamma_{\text{F}} S^{3} .
\label{normalized_scale_002a}
\end{equation}

Equations~(\ref{normalized_scale_001}) and (\ref{normalized_scale_002a}) arise from different modeling assumptions and approximations, and thus cannot be directly compared. Notably, the third term in Eq.(\ref{normalized_scale_001}) originates from the closure scheme in the FP hierarchy and has no explicit analogue in the phenomenological LdG expression. However, a useful connection can still be established by considering the small $S$ expansion of Eq.(\ref{normalized_scale_001}), where $\left(1-S\right)^{\nu} \simeq 1-\nu S$, for $S\ll 1$. Retaining only the leading-order contributions, the FP-based equation becomes a cubic polynomial in $S$, which allows term-by-term comparison with Eq.~(\ref{normalized_scale_002a}).

This comparison yields the following identifications:
\begin{equation}
\alpha_{\text{F}} = 6 \gamma_{1} D_{\text{r}} \left(1- \frac{U}{5} \right);
\beta_{\text{F}} = \frac{4}{7} \gamma_{1}D_{\text{r}} U;
\text{ and }
\gamma_{\text{F}} = \frac{12}{35}\gamma_{1} \nu D_{\text{r}} U .
\label{ldg_coefficients_013}
\end{equation}

Spatial inhomogeneities in $\Qvec$ can be included and the formalism developed in Refs.~\cite{kroeger_j_chem_phys_2007,kroeger_j_nonnewton_fluid_mech_2008} yields the following expression for the elastic constant $L$, using the one-constant approximation of the elastic energy density in \eqref{rLdGenergy},
\begin{equation}
L = \frac{2}{9} \frac{\gamma_{1}D}{S^{2}}
    \left[
    1+\frac{14 U}{15\left(7+5 S -12 S_{4}\right)}
    \right],
    \label{elastic_coeff_001}
\end{equation}
where $D$ is the diffusion coefficient.

\begin{figure}
    \centering
    \includegraphics[width=0.95\linewidth]{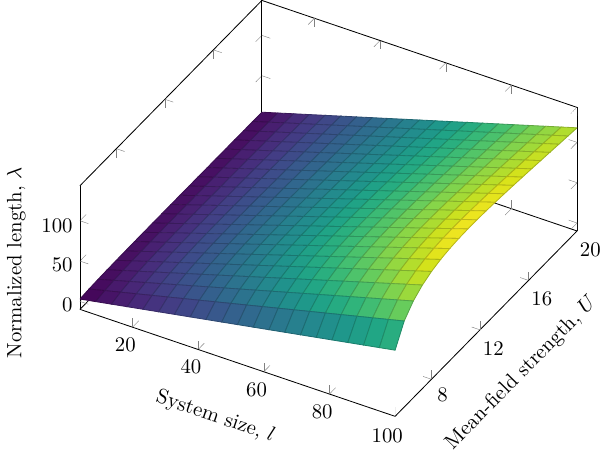}
    \caption{ Normalized length scale $\lambda = l/\xi_{\text{N}}$ estimated from the closure-based derivation of LdG and elastic coefficients, as a function of system size $l$ and interaction strength $U$, in simulation units. The nematic coherence length $\xi_{\text{N}}$ is computed using Eqs.(\ref{ldg_coefficients_013}) and (\ref{elastic_coeff_001}) under the equal elastic constant approximation for the LdG elastic energy density, assuming equal rotational and translational diffusion coefficients for simplicity. The closure function $S_{4}(S) =S\left[1-\left(1-S\right)^{\nu}\right]$ is taken with $\nu=1$, and stationary values of the order parameter are used to ensure consistency with equilibrium conditions.}
    \label{figure_004}
\end{figure}

This expression, together with the relations in Eqs.~(\ref{ldg_coefficients_013}), the solution of the stationary condition $\partial S/\partial t = 0$, and the closure function $S_{4}=S_{4}(S)$, enables us to estimate the nematic coherence length and therefore, the normalized length scale, $\lambda$ in \eqref{normalized_length_001}. 
We perform this estimation using the scalar closure with $\nu =1$ (using the 3D closure relation $S_{4}(S) = S\left[1-\left(1-S\right)^{\nu}\right]$) and constant values of the diffusion coefficients, and summarize the resulting behavior of $\lambda$ as a function of both system size $l$ and interaction strength $U$ in Figure~\ref{figure_004}. This plot illustrates how the characteristic length scale evolves across regimes relevant for our simulations. 

Figure~\ref{figure_004} has some key information about the asymptotic limits of $\lambda$ as a function of $l$ and $U$. Notably, $\lambda$ vanishes for $l \rightarrow 0$ and $U\rightarrow U_{\text{c}}$, where $U_{\text{c}}\simeq 5$ for the 3D MS case. This reflects the divergence of the nematic coherence length at the critical temperature/interaction strength associated with the onset of nematic ordering. As the interaction strength $U$ increases beyond $U_{\text{c}}$, the order parameter saturates, the coherence length stabilizes, and $\lambda$ grows linearly with system size, consistent with expectations for a strongly ordered nematic phase. Importantly, unlike our previous study~\cite{hijar_soft_matter_2024}, where $\lambda$ was estimated semi-empirically by fitting elastic coefficient to simulation data, the current analysis focuses on an analytic derivation of $\lambda$  from a FP framework and a physically grounded closure scheme. This analytic derivation not only validates the asymptotic behavior of $\lambda$ near $U=U_c$ but also provides a robust tool for interpreting confinement effects in mesoscopic nematic simulations.

In the next sections, we use N-MPCD methods to simulate NLCs in square domains with tangent boundary conditions for different values of $l$ (or $R$) and $U$, and use the definition/estimations of $\lambda$ to make connections with previous deterministic LdG studies of the same problem. 
\section{Results}

\subsection{Phase diagrams}
\label{phase_diagram_section}

Equilibrium NLC configurations on square domains with tangent boundary conditions, have been extensively studied as rLdG energy minimizers in a batch of papers \cite{han_siam_2020, yin_phys_rev_lett_2020, robinson_liq_cryst_2017}. Referring to Eq.~\eqref{rLdGenergy}, we can informally identify $\eta$ with $\lambda$ in Eq.~\eqref{normalized_length_001}. The competing NLC equilibria or rLdG energy minimizers are the WORS (with two diagonal defect crosses) for small $\eta$, followed by the diagonal and rotated solutions for sufficiently large $\eta$. The WORS exists as a critical point of the rLdG energy for all $\eta$ but loses stability as $\eta$ increases. The diagonal and the rotated solutions can be distinguished by their vertex profiles; the tangent boundary conditions require the nematic director (leading eigenvector of the rLdG $\Qvec$-tensor order parameter) to be tangent to the square edges and each vertex can be classified as a \emph{splay} or \emph{bend} vertex. The director has a radial profile at a splay vertex whereas the director bends around a bend vertex. The diagonal solution has two diagonally opposite splay vertices and two diagonally opposite bend vertices. The rotated solution has two adjacent splay vertices and two adjacent bend vertices. The unstable critical points can also be classified in terms of the splay or bend character of the square vertices, and the number of interior defects, as will be discussed in the next section.

In this section, we use N-MPCD methods to compute equilibrium/stable NLC configurations for this toy problem, with the three types of mean-field potentials: MS, MG and IK\"O potentials, as described in section~\ref{mean_field_potential_models_section}, as a function of the ordering strength, $U$ and the size of the confinement region, $R$. These are varied in the range $U \in \left[U_{\text{c}},14\right]$, and $a^{-1} R = \left[8,64\right]$, respectively where $U_{\text{c}}\approx 4$ for the MS and MG potentials, $U_{\text{c}}\approx 5.66$ for the IK\"O potential. Four independent simulations, generated by different random numbers series, are conducted for each combination of $U$ and $R$. All the other N-MPCD parameters are fixed with the values given in table~\ref{table_001}. 


\begin{table}[t]
\centering
\small
\begin{tabular}{|>{\raggedright\arraybackslash}p{0.72\columnwidth}|r|}
\hline
Parameter & Value \\
\hline\hline
Mean number of particles per cell, $\bar{N}^{\text{c}}$ & $20$ \\
Time-step for ballistic motion, $\delta t$ (units of $u_{\text{t}}$) & $0.01$ \\
N-MPCD collision time-step, $\Delta t$ (units of $u_{\text{t}}$) & $0.1$ \\
Velocity–orientation coupling parameter, $\chi_{\text{HI}}$ & $0.2$ \\
Tumbling parameter, $\lambda_{\text{t}}$ & $1.5$ \\
Rotational friction coefficient, $\gamma_{\text{R}}$ (units of $u_{\text{t}}^{-1}$) & $0.01$ \\
\hline
\end{tabular}
\caption{Fixed simulation parameters during numerical tests. Parameters $\bar{N}^{\text{c}}$, $\chi_{\text{HI}}$, and $\lambda_{\text{t}}$ are dimensionless.}
\label{table_001}
\end{table}

The numerical tests are initialized with randomly disordered configurations, except when explicitly indicated.  The NLC systems are allowed to thermalize over $6\times 10^{4}$ collision steps. The thermalization stage is followed by a measurement stage encompassing $3\times 10^{4}$ further collision steps. The numerical results in this section correspond to averages over the measurement stage.

We identify topological defects in N-MPCD simulations as localised regions of small order parameter,  $S^{\text{c}}\sim 0.5$. Although the order parameter is theoretically expected to vanish in the strongly distorted regions near defect cores~\cite{chaikin_principles_1995}, such small values are not observed in our simulations. This discrepancy arises because the N-MPCD scalar order parameter is computed at the level of collision cells---the smallest spatial units on which fields are defined---each containing a relatively small number of particles. For these small ensembles, typically of size $\bar{N}^{\text{c}}\sim20$, the eigenvalues of $\matriz{Q}^{\text{c}}$ exhibit finite-size effects. In particular, the largest eigenvalue remains noticeably greater than zero in disordered regions~\cite{eppenga_mol_phys_1984}.

\begin{figure}
    \centering
    \includegraphics[width=0.95\linewidth]{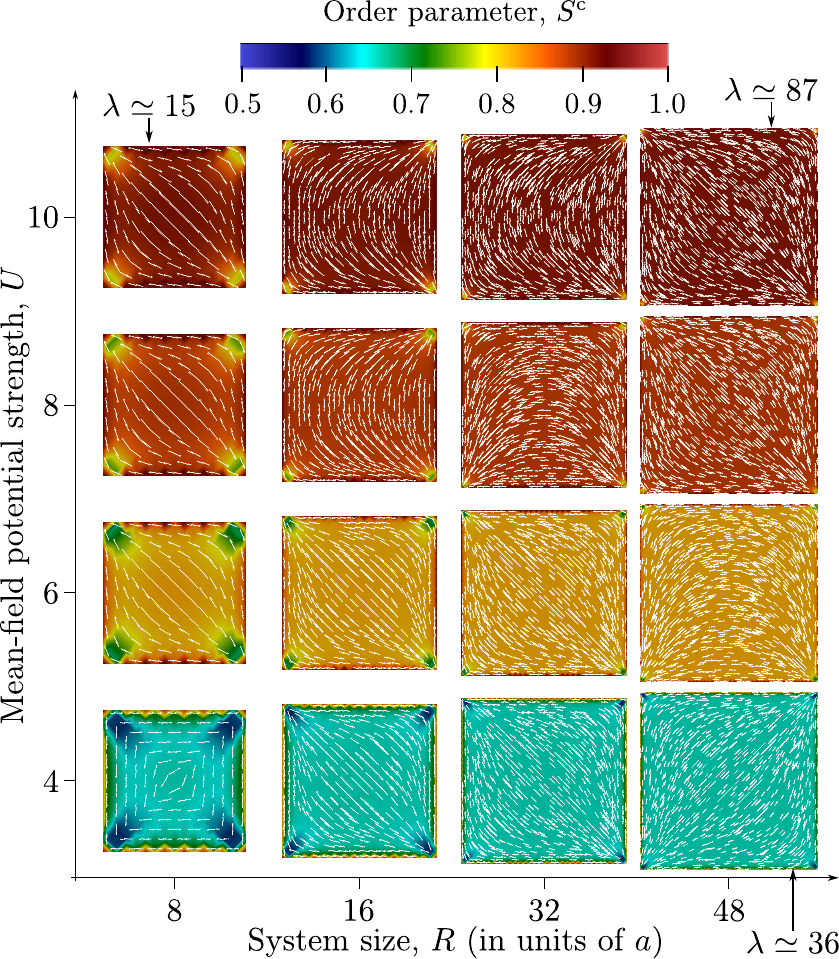}
    \caption{Average equilibrium configurations in square domains of different sizes for N-MPCD fluids, simulated through MS interactions of different strength, $U$. The color bar defines $S^{\text{c}}$, as dictated by the bar at the top. The small interior lines label the local director, $\unitvc{n}^{\text{c}}$. Relative system sizes are not realistic but chosen to ease readability and interpretability of the figures.}
    \label{figure_005}
\end{figure}

\begin{figure}
    \centering
    \includegraphics[width=0.95\linewidth]{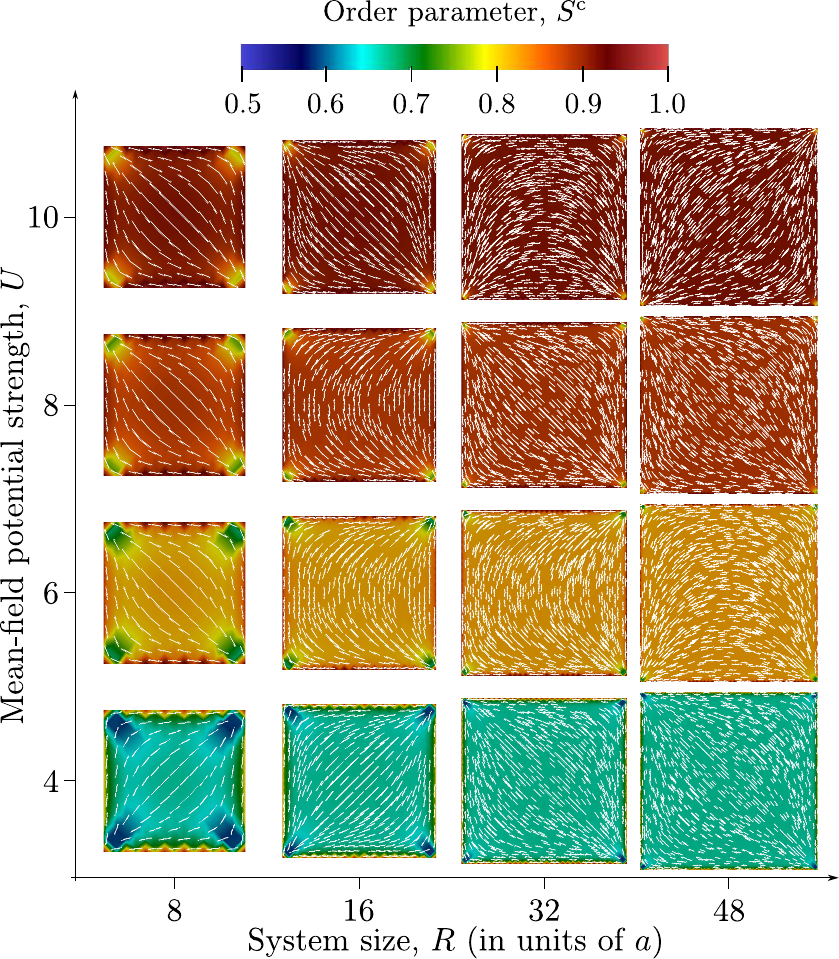}
    \caption{MG equilibrium structures with N-MPCD methods; the color bar and the interior lines have the same interpretation as in figure~\ref{figure_005}.}
    \label{figure_006}
\end{figure}

\begin{figure}
    \centering
    \includegraphics[width=0.95\linewidth]{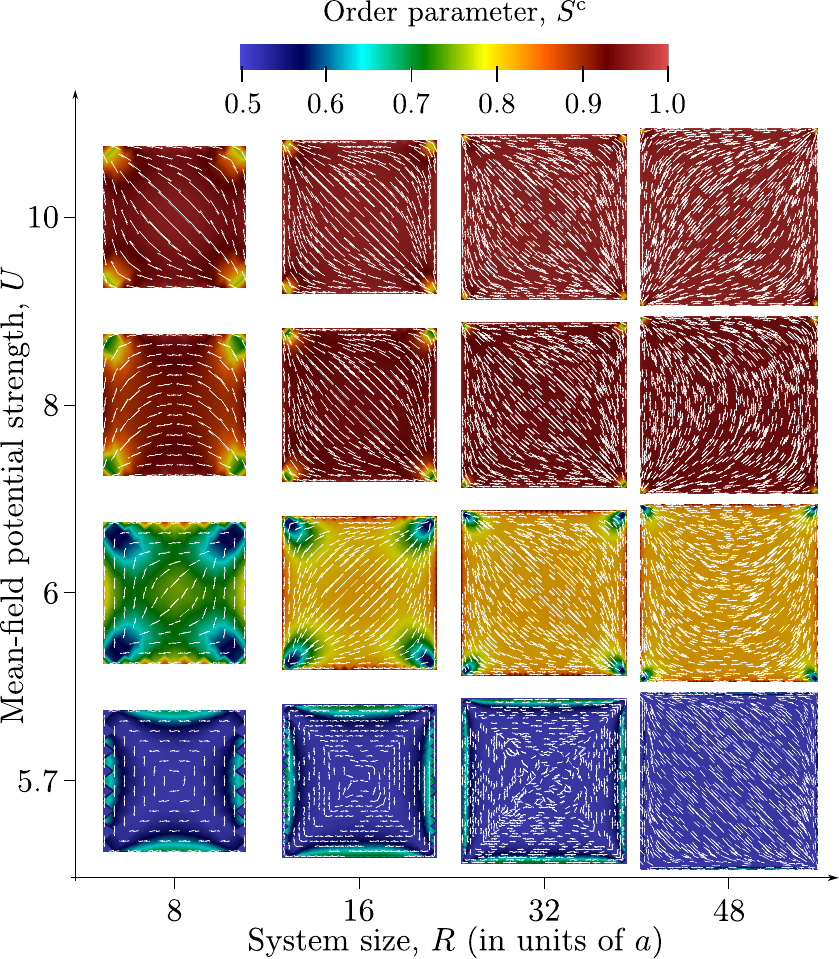}
    \caption{ IK\"O equilibrium structures simulated with N-MPCD methods; the color bar and the interior lines have the same interpretation as in figure~\ref{figure_005}.}
    \label{figure_007}
\end{figure}

For large $U$ and $R$ ($\lambda \gtrsim 15$), all systems equilibrate either to the diagonal or the rotated solution, independently of the underlying mean-field interaction. Both solutions have two splay-type and two bend-type square vertices. These solutions are illustrated in Figs.~\ref{figure_005}, \ref{figure_006}, and \ref{figure_007} for the MS, MG, and IK\"O cases, respectively. For reference, figure~\ref{figure_005} includes the estimations of $\lambda$ for some specific systems. 


The MS and MG equilibrium configurations are essentially indistinguishable for $U > 4$ and $R>8\,a$, as can be noticed by comparing Figs.~\ref{figure_005} and \ref{figure_006}. This suggests that for these simulation parameters, the spatial variations of $\matriz{Q}^{\text{c}}$, as accounted for the MG potential in Eq.~(\ref{model_004}), do not significantly affect $S$ or the admissible director patterns (diagonal or rotated) and the MS/MG predictions are wholly consistent with the continuum rLdG predictions.  

However, the MS and MG equilibrium configurations differ  in small domains, $R=8\,a$, close to the critical anchoring strength, $U_{\text{c}}=4$. The MS configuration resembles the WORS solution, consistent with the unique rLdG energy minimiser for small domains and high temperatures (corresponding to small values of $U$). In fact, the WORS is the unique MS equilibrium configuration on square domains with tangent boundary conditions for small values of $R$ and  $U\simeq U_{\text{c}}$, \text{i.e.}, as $\lambda \rightarrow 0$~\cite{hijar_soft_matter_2024}. However, the MG equilibrium configuration has a more diagonal profile for $R=8\,a$ and $U_{\text{c}}=4$. Informally speaking, $\Delta Q_{ij} = 0$ for the WORS solution whereas $\Delta Q_{ij}\propto Q_{ij}$ for the diagonal solution (see \cite{han_siam_2020}). Thus the elastic effects of the diagonal solution compete with the main part of the MS potential in the MG potential \eqref{model_003} and the resulting mean-field MG energy minimizer is the diagonal-type solution for small $R$ and for $U \sim U_{\text{c}}$. It is not clear if the MG potential can recover the WORS solution in any parameter regime. 

In comparison with the MS and MG equilibrium configurations, the IK\"O equilibrium configurations are characterized by larger values of $S^{\text{c}}$ in the central region of the square and defects that extend over larger distances from the vertices, as shown in Fig.~\ref{figure_007}. These effects are produced by the non-linear dependence of the mean-field potential on the order parameter, which is ultimately expressed in the argument of the exponential function in the probability distribution function in Eq.~(\ref{simulations_005}). 
The IK\"O potential is effectively equivalent to a MS potential with a $S^{\text{c}}$-dependent interaction strength $U_{\text{mf}}(S)$, 
satisfying $U_{\text{mf}} \gg U$ for $S^{\text{c}} \simeq 1$; this explains the large amount of order at the center of the squares. In contrast, $U_{\text{mf}} < U$ for $S^{\text{c}} \simeq 0$. 
This qualitatively explains why IK\"O nematic defects are noticeably larger than those reported for the MS and MG equilibrium configurations. In Fig.~\ref{figure_007}, the most dramatic example is observed for $U = 5.7 \simeq U_{\text{c}}$ and $R = 8\, a$ where the orientation pattern is similar to the WORS solution and topological deffects extend over distances comparable to the size of the simulated system. Further, for $U\sim U_c = 5.7$, the IK\"O equilibrium configuration resembles the WORS solution not only for for $R=8\,a$, but also for moderately large, $R=16\,a$, domains. Within larger squares, $R=32\,a$, the average director pattern is the WORS solution near the edges, surrounding a disordered central region without a well-defined director, and the familiar diagonal solution is recovered for $R=48\,a$.
These results illustrate that the nonlinear dependence of the IK\"O potential on the order parameter leads to distinct defect structures and smaller ordered regions compared to MS and MG interactions, emphasizing its relevance in scenarios where strong nonlinear effects are anticipated.
 
\subsection{Energy pathways and unstable configuration}

In \cite{hijar_soft_matter_2024}, the authors study the relaxation of N-MPCD systems from random disordered configurations to ordered NLC configurations, with the MS mean-field potential. The relaxation pathways exhibit transient metastable structures with complex interior defect constellations, and identify some universal patterns and defect annihilation laws. 
In this section, we replicate the work in \cite{hijar_soft_matter_2024} and study relaxation pathways with the MS, MG and IK\"O potentials respectively.

The relaxation pathway is defined by the time evolution of the elastic energy; the energy decreases with time along a relaxation pathway to an equilibrium configuration. The total elastic energy per unit length, as stored in a 2D nematic texture, is defined to be ~\cite{de_gennes_1993,pearce_soft_matter_2021}
\begin{equation}
E\left(t\right) 
= \frac{L}{2} 
\int dx^{2} \frac{\partial Q_{\alpha\beta}(t)}{\partial x_{\gamma}}   
            \frac{\partial Q_{\alpha\beta}(t)}{\partial x_{\gamma}},
\label{energy_pathways_001}
\end{equation}
where $L$ is the elastic constant for the one-constant elastic energy density and the integral is performed over the square domain. 

The elastic energy is approximated in the N-MPCD framework as \begin{equation}
E\left(t_{i}\right) 
= \frac{L}{2}  a^{2} 
   \sum_{\text{cells}} \frac{\partial Q^{\text{c}}_{\alpha\beta}(t_{i})}{\partial x_{\gamma}}   
                       \frac{\partial Q^{\text{c}}_{\alpha\beta}(t_{i})}{\partial x_{\gamma}} ,
\label{energy_pathways_002}
\end{equation}
where $\sum_{\text{cells}}$ extends over all collision cells and derivatives are approximated by discrete differences, \textit{e.g.},
\begin{equation}
\frac{\partial Q^{\text{c}}_{\alpha\beta}(t_{i})}{\partial x_{\gamma}} 
\simeq
\frac{Q^{\text{c}}_{\alpha\beta}\left(x_{\gamma} + a,t_{i}\right) - Q^{\text{c}}_{\alpha\beta}\left(x_{\gamma} - a,t_{i}\right)}{2 a},
\label{energy_patways_003}
\end{equation}
where $t_{i}$ denotes the time of the multi-particle collision event.

\begin{figure}
    \centering
    \includegraphics[width=0.95\linewidth]{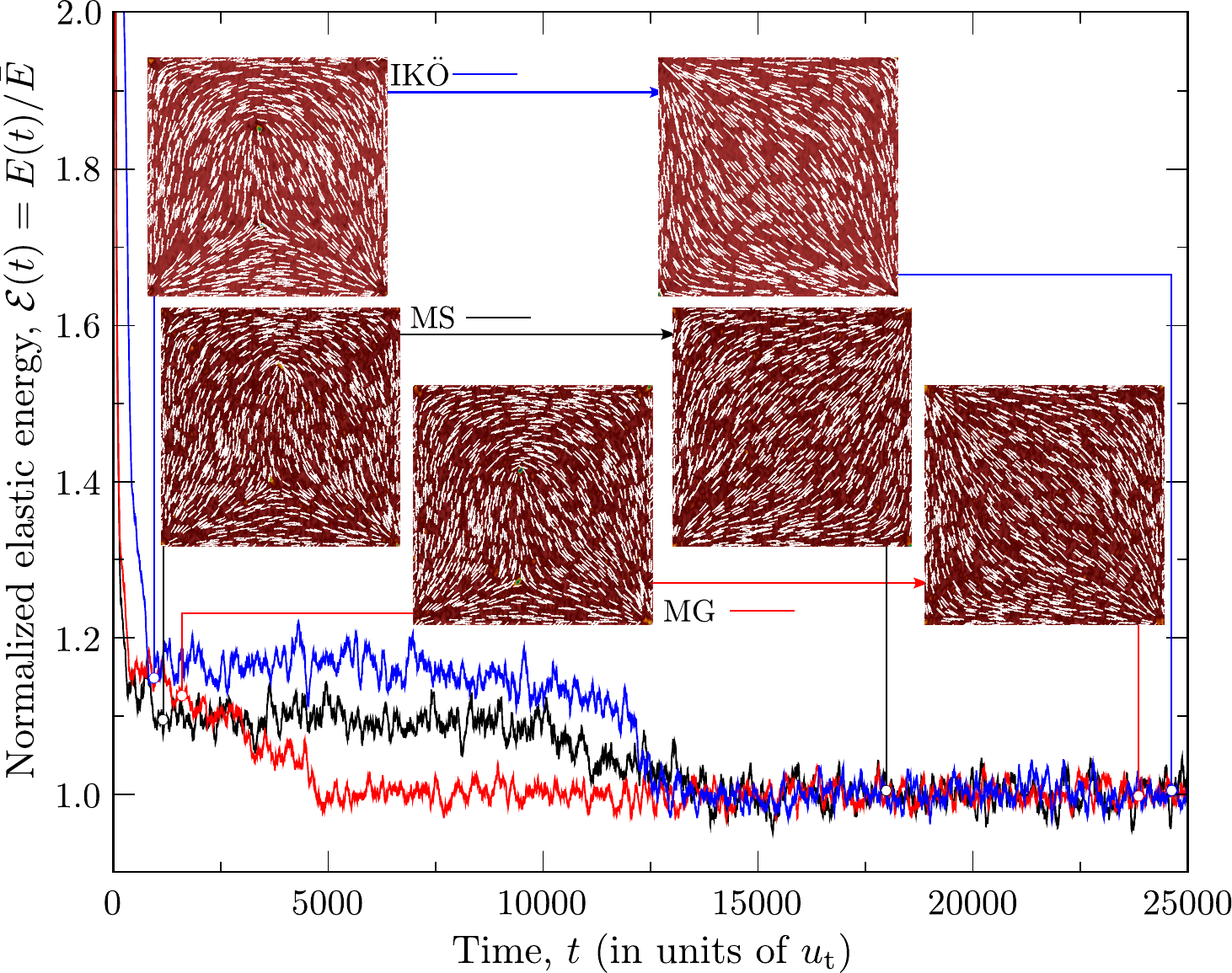}
    \caption{Energy relaxation to equilibrium in N-MPCD simulations guided by the MS potential (black curve), the MG potential (red curve), and the IK\"O potential (blue curve). Insets indicate the director field pattern at selected times during an energy plateau (left) and in equilibrium (right).}
    \label{figure_008}
\end{figure}

The elastic energy is normalized with respect to the elastic energy of the diagonal solution, which is calculated by the time average of $E(t_{i})$ over the measurement stage:
\begin{equation}
\bar{E} = \frac{1}{n_{\text{m}}}\sum_{i=1}^{n_{\text{m}}}E\left(t_{i}\right),
\label{energy_pathways_004}
\end{equation}
where $n_{\text{m}}$ denotes the total number of collision events after thermalization.
The time-series of normalized energies, $\mathcal{E}(t)=E(t)/\bar{E}$, are noisy curves exhibiting a fast initial decay followed by cascades wherein the energy is quasi-stationary or slowly decreasing, until equilibrium is reached. The cascades signal the approach to a metastable configuration, that could be identified with an unstable saddle point of the rLdG energy; see~\cite{robinson_liq_cryst_2017}. This is illustrated in Fig.~\ref{figure_008} for three N-MPCD systems of size $R=64\,a$, simulated with the MS, the MG, and the IK\"O potentials and $U = 10$, ($\lambda \simeq 117$). In this case, the final equilibrium configuration is the diagonal solution for all three curves. There is a semi-stationary unstable configuration, that resembles a rotated solution (with two adjacent splay vertices) and an interior pair of $\pm \frac{1}{2}$ defects roughly located near the square centre. The normalized energies of the unstable configurations are similar in the MS and MG cases, and they are about $20$\% smaller than the IK\"O case. This can be partly explained on the grounds that defects are energetically more expensive in the IK\"O framework.

\begin{figure}
    \centering
    \includegraphics[width=0.95\linewidth]{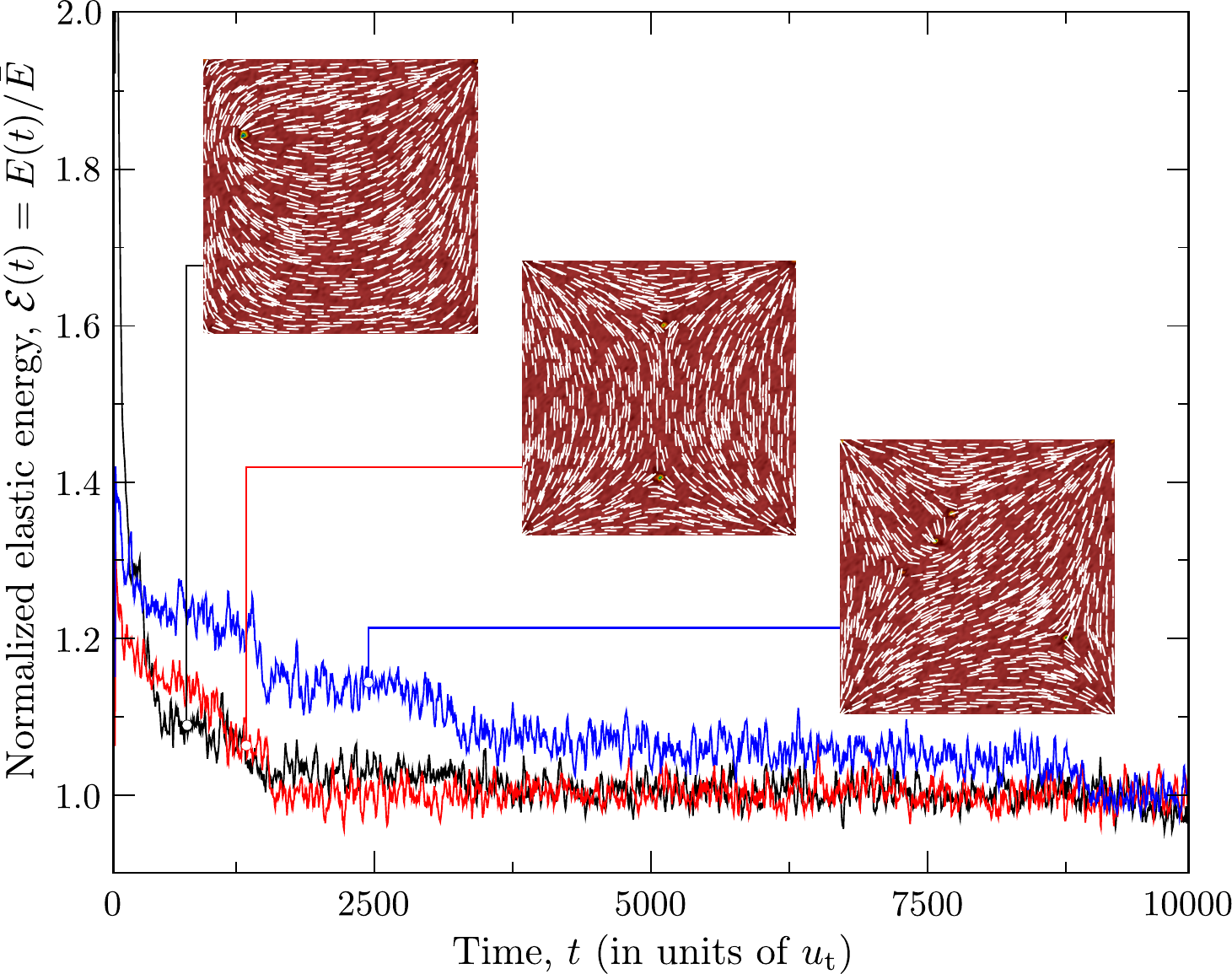}
    \caption{ Three different N-MPCD systems simulated with the IK\"O mean-field interaction with $\lambda \simeq 51$; also see Figure~\ref{figure_007}.}
    \label{figure_009}
\end{figure}

In Fig.~\ref{figure_009}, we illustrate N-MPCD relaxation pathways with the IK\"O model in systems with $U=14$ and $R=32\, a$, ($\lambda \simeq 51$). For the black noisy curve, the pattern has a single $+1/2$ defect located close to a bend vertex. This is qualitatively similar to the index-$1$ $J\pm$ saddle point of the rLdG energy reported in~\cite{yin_phys_rev_lett_2020}, for relatively large domain sizes. This metastable configuration has three bend vertices and one splay vertex, and the $1/2$ interior defect balances the overall topological charge. This pattern has also been predicted as an unstable transient state in other independent investigations~\cite{robinson_liq_cryst_2017}. In the remaining two cases, simulations are initialized with orientations distributed radially according to $\unitvc{u}_{i}=\unitvc{e}_{r}\left( \mathbf{r}_{i} \right)$, where $\unitvc{e}_{r}\left( \mathbf{r}_{i} \right)$ is the unit radial vector in polar coordinates at position $\mathbf{r}_{i}$ (instead of random initial orientations). Generally speaking, we observe that some unstable structures appear more frequently for some specific initial conditions, although we do not attempt to establish a relationship between initial configurations and unstable transient states, which is outside the scope of our study. 

In Fig.~\ref{figure_009}, we observe two further unstable transient states for which $\mathcal{E}(t)$ has a shallow plateau, with the initial radial orientation distributions. The red noisy curve exhibits an unstable transient state with two symmetric interior $-1/2$ defects and four splay vertices. This unstable configuration has been also observed in rLdG settings by Robinson \textit{et al}.~\cite{robinson_liq_cryst_2017}, when numerically computing critical points of the rLdG free energy in \eqref{rLdGenergy}. The blue noisy curve in Fig.~\ref{figure_009} exhibits another unstable transient configuration with four splay vertices and a pair of interior $\pm 1/2$ defects located symmetrically around the square diagonal. This solution has some similarity with the index-$3$ $T$ or $T\pm$ saddle points of the rLdG energy reported in~\cite{yin_phys_rev_lett_2020}. 
These comparisons are interesting but can be made more precise and quantitative by considering different system sizes, interaction strengths, initial conditions supplemented with alternative methods to calculate the energy and stability of the proposed configurations as those used in, \textit{e.g.}, reference~\cite{robinson_liq_cryst_2017}. Of course, one should also use map the values of $l$ and $U$ to the reference values for the rLdG studies in~\cite{yin_phys_rev_lett_2020} and \cite{han_siam_2020, robinson_liq_cryst_2017}.

\begin{figure}
    \centering
    \includegraphics[width=1.0\linewidth]{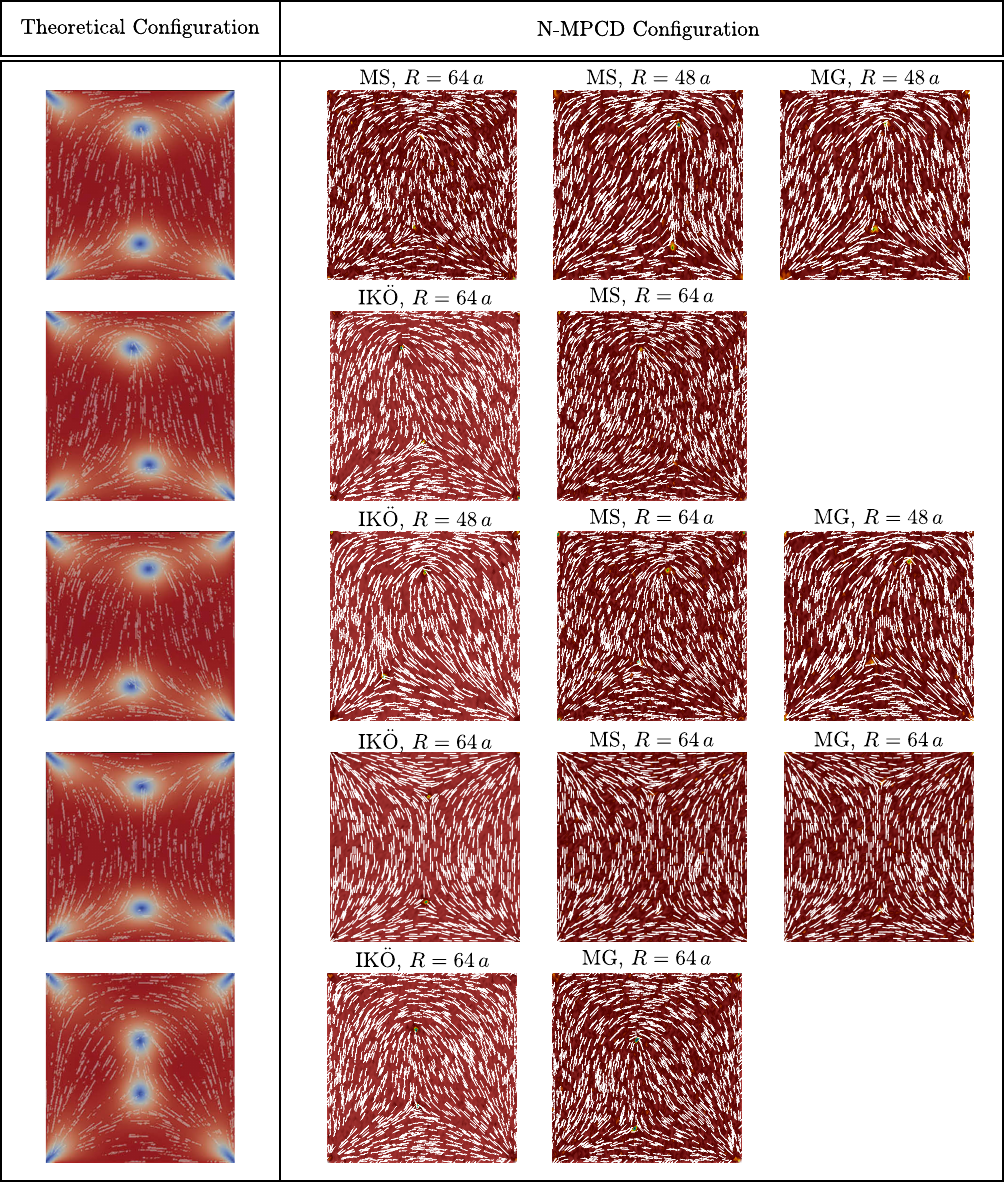}
    \caption{Comparison between unstable equilibrium configurations obtained theoretically in reference~\cite{robinson_liq_cryst_2017} and by means of N-MPCD simulations with $U=10$. The label at the top of the insets indicate the type of mean-field potential (MS, MG, or IK\"O) used during simulations and the size of the system.}
    \label{figure_010}
\end{figure}

In~\cite{robinson_liq_cryst_2017}, the authors use numerical deflation methods to identify $19$ unstable critical points of the rLdG free energy on square domains with tangent boundary conditions. In Fig.~\ref{figure_010}, we recover $9$ of these unstable configurations by means of N-MPCD simulations guided by MS, MG, or IK\"O mean-field potentials respectively. For convenience, we compare the rLdG predictions from \cite{robinson_liq_cryst_2017} with the N-MPCD results in two separate figures. These unstable configurations have two defects located close to the short symmetry axis of the square. The unstable configuration has a rotated profile with an interior $\pm 1/2$ defect pair, except for the fourth row wherein the unstable configuration has two  aligned $-1/2$ defects and four splay vertices. These transient solutions have also been observed in the energy relaxation pathways. We note that this transient states are qualitatively similar to index-$2$ saddle points of the rLdG energy reported in \cite{yin_phys_rev_lett_2020} for certain system sizes. 

\begin{figure}
    \centering
    \includegraphics[width=1.0\linewidth]{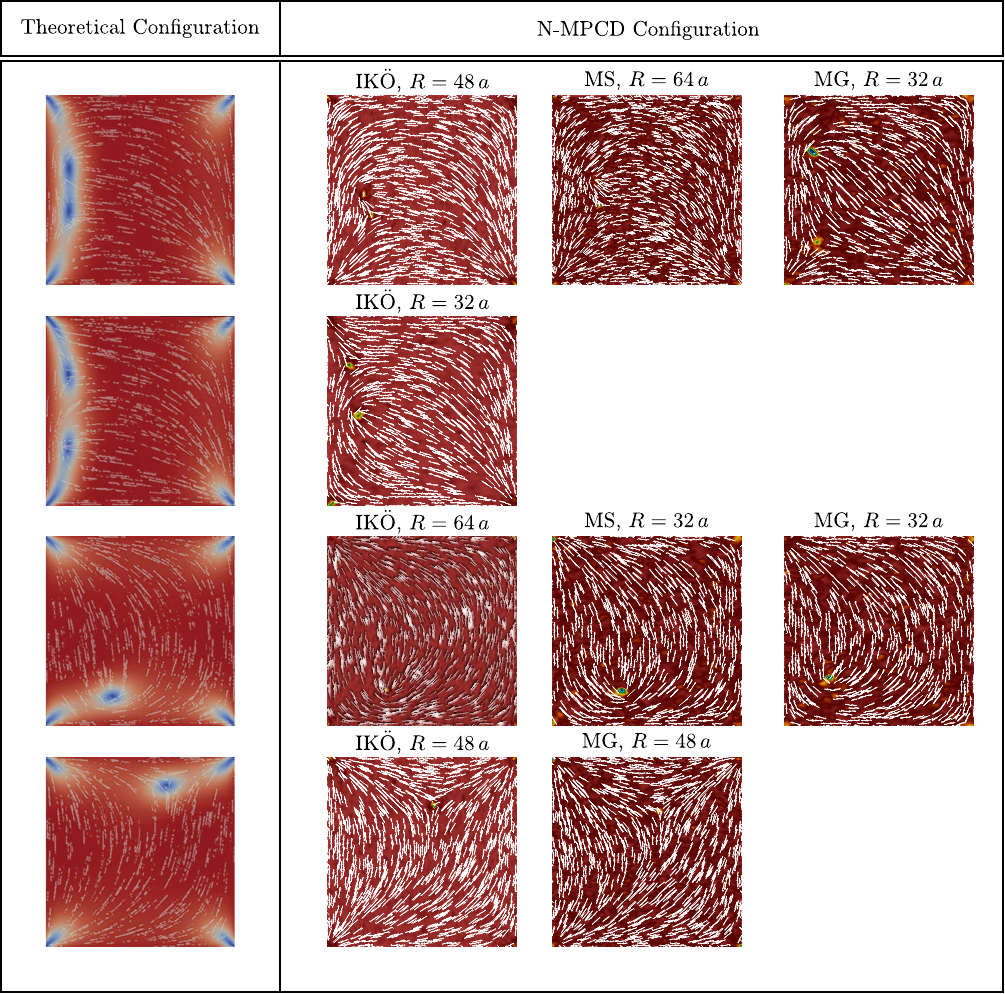}
    \caption{Similar to figure~\ref{figure_010} for miscellaneous unstable equilibrium configurations having one or two interior defects.}
    \label{figure_011}
\end{figure}

In Fig.~\ref{figure_011}, we plot miscellaneous structures with one interior defect or two interior defects localised near a square edge. The numerical results in Figs.~\ref{figure_010} and \ref{figure_011} are obtained with $U=10$. The inset in each case labels the mean-field potential and the system size under consideration. The configurations with one interior defect (localised near a vertex) are reminiscent of the index-$1$ $J\pm$ rLdG saddle points reported in \cite{yin_phys_rev_lett_2020} whereas the top two rows are reminiscent of the index-$2$ $J$ rLdG saddle point reported in \cite{yin_phys_rev_lett_2020}. The numerical experiments are relatively sparse to make concrete connections but suggest universal patterns in the multiplicity and locations of interior defects, independent of underlying mean-field models.

We observe unstable transient states with all three mean-field potentials and for all ranges of system sizes and interaction strength under consideration, suggesting that such transient states with symmetric interior defects are ubiquitous and can play a crucial role in nematodynamics. The MS, MG, and IK\"O configurations are qualitatively similar and independent of the finer differences between the mean-field potentials. Equally importantly, these unstable transient states have rLdG counterparts as unstable saddle points of the rLdG free energy in \eqref{rLdGenergy}, implying that the continuum rLdG theory can successfully capture stable and unstable NLC configurations for large system sizes and low temperatures (large values of $U$), regardless of the underlying microscale physics. It is also remarkable that we observe high-index unstable saddle points of the rLdG free energy in the relaxation pathways induced by the MS, MG and IK\"O potentials, suggesting that not only index-$1$ but also higher index saddle points can be observable in nematodynamics or physically relevant transition pathways in NLC systems. Altogether, these results affirm the robustness of the continuum LdG framework for describing equilibrium and non-equilibrium phenomena for large systems in the well-ordered nematic regime. Data of some examples illustrating the relaxation dynamics studied in this section are available at~\cite{hijar_mendely_data_2026}.  

\section{Conclusions}
\label{conclusions_section}
This paper can be viewed as a sequel to \cite{hijar_soft_matter_2024} wherein the authors conduct detailed N-MPCD simulations of 2D polygons with tangent boundary conditions, employing the standard Maier-Saupe (MS) mean-field potential. The authors conclude that the MS-based N-MPCD simulations are largely consistent with the continuum rLdG predictions in \cite{han_siam_2020}, for equilibrium and metastable NLC configurations in polygons, for sufficiently large system sizes and well ordered systems (characterized by relatively large values of $U$ or low temperatures).

In this paper, we focus on three questions for a representative and well-studied model problem: (i) do the N-MPCD predictions depend on the choice of the underlying mean-field potential; (ii) do the N-MPCD predictions agree with the continuum rLdG predictions and if so, how does the agreement depend on the choice of the mean-field potential and the closure approximations and (iii) can N-MPCD reveal new phenomena outside the scope of continuum theories, perhaps by tailoring the mean-field potentials etc.?

Our simulations suggest that for sufficiently large $R$ and $U$, \textit{i.e.}, large and ordered systems, the N-MPCD predictions appear to be independent of the details of the collision operators, \textit{i.e.}, the choice of the underlying mean-field potentials. All three potentials under consideration, the MS, MG and IK\"O, yield the diagonal and rotated solutions in the ordered macroscopic regime. Further, the choice of the mean-field potential and the closure approximations determine the critical interaction strength, $U_{\text{c}}(MF)$, and the nematic coherence length, $\xi_{\text{N}} (MF)$, in 2D and 3D. As such, these estimates are different for different mean-field potentials, although we only compute $\xi_{\text{N}}(MS)$ in 3D, in this paper. However, our results suggest that the N-MPCD simulations are almost wholly consistent with the continuum rLdG predictions for $U\gg U_{\text{c}}(MF)$ and for $R \gg \xi_{\text{N}}(MF)$. These bounds also define the regime of validity of the rLdG theory. It is remarkable that the N-MPCD simulations reproduce similar equilibrium and unstable transient configurations (identified as unstable saddle points of the rLdG free energy) as the continuum rLdG theory, for $U\gg U_{\text{c}}(MF)$ and for $R \gg \xi_{\text{N}}(MF)$, and the unstable transient configurations are universally observed regardless of the choice of the underlying mean-field potential. The equilibrium configurations have a characteristic defect-free interior director profile whereas the metastable/transient configurations have interior fractional $\pm 1/2$ or $\pm 1$ nematic defects, or even line defects along the diagonal or near the square edges. There seem to be universal topological rules for the locations and multiplicity of the interior defects, \textit{e.g.}, a $\pm 1/2 $ interior defect pair is accompanied by two splay vertices and two bend vertices; a pair of symmetric $-1/2$ nematic defects is accompanied by four splay vertices; a single interior $1/2$ defect is usually localised near a bend vertex and a single interior $-1/2$ defect is usually localised near a splay vertex, and these rules are universal.

Finally, the particle-based stochastic N-MPCD simulations do not reveal new equilibrium or non-equilibrium structures across the MS, MG and IK\"O potentials, for $U\gg U_{\text{c}}(MF)$ and for $R \gg \xi_{\text{N}}(MF)$. One might then argue that phenomenological models, of which the rLdG or LdG theory is a hugely successful example, with the correct order parameters dictated by molecular symmetries and the phenomenological coefficients fitted to empirical data, are perfectly adequate for large well-ordered systems. However, it is of fundamental scientific interest to understand how finer microscopic or mesoscopic details propagate to macroscopic length scales, where are such details important and can we amplify these details to control and manipulate soft matter systems? Whilst our work affirms the robustness of the continuum LdG theory, which is of independent scientific importance, future work will focus on more detailed studies of N-MPCD simulations of confined nematics and the use of such simulations in multiscale and multiphysics liquid crystal theories.



\section*{Conflicts of interest}
There are no conflicts to declare.

\section*{Acknowledgments}

HH thanks La Salle University Mexico for financial support under grant NEC 22-23. AM is supported by a Leverhulme Research Project Grant RPG-2021-401 and a Leverhulme International Academic Fellowship IAF-2019-009. Computational resources were provided by the HiPer Computing iLab at the Research Center of La Salle University Mexico.

\appendix

\section{Derivation of the closure relation Eq.~(\ref{ldg_coefficients_006})}
\label{appendix_001}

To relate $\langle u_{i} u_{j} u_{k} u_{l} \rangle$ with $\langle u_{i} u_{j} \rangle$, we consider the definition of the Cartesian symmetric traceless tensors of rank $2$ and $4$, namely~\cite{kroeger_j_nonnewton_fluid_mech_2008},
\begin{equation}
A^{[2]}_{ij} = \left\langle u_{i} u_{j} \right\rangle -\frac{1}{2} \delta_{ij}
                 , 
\label{details_appendix_001}
\end{equation}
and
\begin{equation}
A^{[4]}_{ijkl} = \left\langle u_{i} u_{j} u_{k} u_{l} \right\rangle
- \left\langle \unitvc{u} \unitvc{u} \matriz{I} \right\rangle^{\text{sym}}_{ijkl}
+\frac{1}{8} \left( \matriz{I} \matriz{I} \right)^{\text{sym}}_{ijkl} ,
\label{details_appendix_002}
\end{equation}
respectively, where the superscript $^{\text{sym}}$ denotes the symmetric normalized part of tensors defined as
\begin{eqnarray}
\left( \matriz{X}\matriz{Y} \right)^{\text{sym}}_{ijkl}
& = &  \frac{1}{6}
\left(
  X_{ij}Y_{kl}
+ X_{ik}Y_{jl}
+ X_{il}Y_{jk}  \right. \nonumber \\
&   & \left.
+  X_{jk} Y_{il}
+ X_{jl} Y_{ik}
+ X_{kl} Y_{ij}
\right).
\label{details_appendix_005}
\end{eqnarray}


$A^{[2]}_{ij}$ and $A^{[4]}_{ijkl}$ are also related with $n_{i}$, $S$, and $S_{4}$ through~\cite{kroeger_j_nonnewton_fluid_mech_2008}
\begin{equation}
A^{[2]}_{ij} = S \left( n_{i} n_{j} -\frac{1}{2} \delta_{ij} \right),
\label{details_appendix_003}
\end{equation}
and 
\begin{equation}
A^{[4]}_{ijkl} = S_{4} \left[ n_{i}n_{j}n_{k}n_{l} 
- \left( \unitvc{n} \unitvc{n} \matriz{I} \right)^{\text{sym}}_{ijkl}
+\frac{1}{8} \left( \matriz{I} \matriz{I} \right)^{\text{sym}}_{ijkl}\right].
\label{details_appendix_004}
\end{equation}

From Eqs.~(\ref{details_appendix_001}) to (\ref{details_appendix_004}) we obtain 
\begin{eqnarray}
\langle u_{i}u_{j}u_{k}u_{l}\rangle & = & S_{4} n_{i}n_{j}n_{k}n_{l}
+ \left(S_{2}-S_{4}\right) \left(\unitvc{n}\unitvc{n}\matriz{I} \right)^{\text{sym}}_{ijkl} \nonumber \\
&  & +\frac{1}{8}\left(3-4S+S_{4}\right) \left(\matriz{I}\matriz{I}\right)^{\text{sym}}_{ijkl}.
    \label{details_appendix_006}
\end{eqnarray}

By expanding the symmetric tensor $\left(\unitvc{n}\unitvc{n}\matriz{I} \right)^{\text{sym}}$, using
\begin{equation}
n_{i}n_{j} = \frac{1}{S} \langle u_{i} u_{j} \rangle -\frac{1}{2 S} \left(1-S\right)\delta_{ij},
    \label{details_appendix_007}
\end{equation}
and simplifying, we finally obtain Eq.~(\ref{ldg_coefficients_006}) with the coefficients $\alpha$, $\beta$, and $\gamma$ defined in Eq.~(\ref{ldg_coefficients_007}).


\bibliography{liquid_crystals}

\bibliographystyle{rsc} 

\end{document}